\documentstyle[aps,epsf,bbold]{revtex}
\newcommand {\beq}
            {\begin{equation}}
\newcommand {\beqa}
            {\begin{eqnarray}}
\newcommand {\eeq}
            {\end{equation}}
\newcommand {\eeqa}
            {\end{eqnarray}}
\begin {document}
\begin{title}
{
\hfill{\small {\bf MKPH-T-99-16}}\\
{\bf Invariant amplitudes for coherent electromagnetic pseudoscalar 
production from a spin-one target (II): crossing, multipoles and observables}
\footnote{Supported by the Deutsche 
Forschungsgemeinschaft (SFB 443)}}
\end{title}
\bigskip
\author{
Hartmuth Arenh\"ovel}
\address{
Institut f\"ur Kernphysik, 
Johannes Gutenberg-Universit\"at,
D-55099 Mainz, Germany}
\maketitle
\begin{abstract}
The formal properties of the recently derived set of linearly 
independent invariant amplitudes for the 
electromagnetic production of a pseudoscalar particle from a spin-one 
particle have been further exploited. The crossing properties are 
discussed in detail. Since not all of the amplitudes have simple crossing 
behaviour, we introduce an alternative set of basic amplitudes 
which are either symmetric or antisymmetric
under crossing. The multipole decomposition is given, and 
the representation of the multipoles as integrals over the invariant
functions weighted with Legendre polynomials is derived. 
Furthermore, differential cross section and 
polarization observables are expressed in terms of the 
corresponding invariant functions.  
\end{abstract}

\section{Introduction}\label{introduction}
Recently we have derived a set of 13 linearly independent invariant 
amplitudes for the electromagnetic production of a pseudoscalar particle 
from a spin-one particle in \cite{Are98}. 
These amplitudes respect Lorentz and gauge invariance and the 
$T$-matrix can be represented by a linear superposition of them
with invariant functions as coefficients, which depend on the Mandelstam 
variables only. Nine of these amplitudes are purely transverse and 
describe photoproduction while the remaining four appear in 
electroproduction in addition, describing charge and longitudinal current 
contributions. 

In the present work we want to elaborate this formalism in further detail
by exploring, for example, the crossing properties of the amplitudes and 
corresponding invariant functions and deriving explicit expressions 
for the observables.
To this end, we first review briefly in Sect.\ \ref{review} the set of basic 
amplitudes and discuss then in Sect.\ \ref{crossing} their crossing 
properties. In view of the fact that not all 13 basic amplitudes 
introduced in \cite{Are98} have simple crossing properties, we propose 
another alternative set of amplitudes which are either symmetric or 
antisymmetric under crossing. As next the multipole
decomposition of the $T$-matrix will be discussed in Sect.\ 
\ref{multipole} where we give explicit expressions of the multipoles as 
integrals over the invariant functions weighted with linear combinations 
of Legendre polynomials. Then, in Sect.\ \ref{observable} we will define 
the various observables, i.e., differential cross section, beam and target 
asymmetries and the final state polarization observables. They are 
represented by structure functions which can be expressed in terms of 
the invariant functions.

\section{Brief review of the invariant amplitudes}\label{review}

\begin{figure}
\centerline{%
\epsfxsize=8.0cm
\epsffile{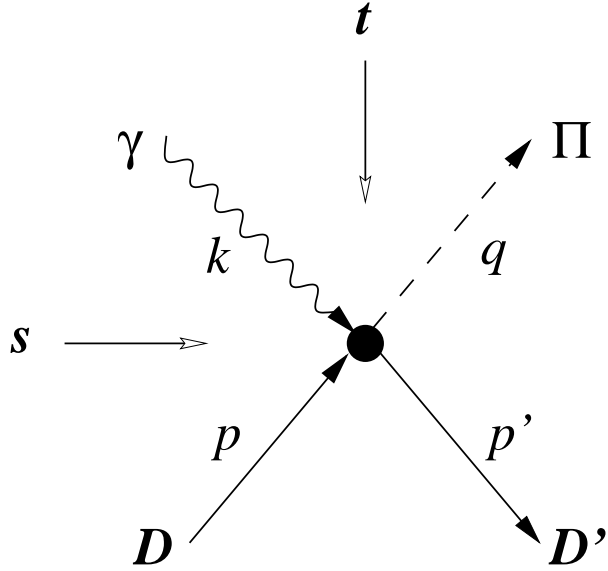}
}
\vspace*{1cm}
\caption{Diagram for coherent electromagnetic production of a pseudoscalar 
particle $\Pi$ on a spin-one particle $D$. The arrows labeled $s$ and $t$ 
indicate $s$- and $t$-channels.
\label{fig1}
}
\end{figure}

As in \cite{Are98}, we consider as basic reaction the coherent pseudoscalar 
production by photoabsorption or electron scattering from a spin-one target 
\begin{eqnarray}
\gamma^{(*)}(k) + D(p) \rightarrow D(p') + \Pi (q)\,, 
\end{eqnarray}
where the real or virtual photon has momentum $k$ and polarization vector 
$\epsilon$, $D$ is a spin-one particle of mass $M$, henceforth called 
$D$ for brevity, with initial and final momenta $p$ and $p'$ and 
$(S=1)$-spinors $U(\vec p,m)$ and $U(\vec p^{\,\prime},m')$, respectively, 
and $\Pi$ a pseudoscalar meson of mass $m$, henceforth called $\Pi$, 
with momentum $q$ (see Fig.\ \ref{fig1}). 

As a first step we had derived in \cite{Are98} using the principles 
of Lorentz covariance and parity conservation a set of basic amplitudes 
for the representation of the invariant reaction matrix element 
${\cal M}_{fi}$. They are listed  
in Table \ref{tab0}, where we use as a shorthand $U'$ for 
$U^\dagger(\vec p^{\,\prime},m')$ and $U$ for $U(\vec p,m)$. Furthermore, 
we have introduced for convenience a 
covariant pseudoscalar by contraction of the four-dimensional 
Levi-Civita tensor $\varepsilon^{\mu\nu\rho\sigma}$ with four Lorentz 
vectors $a,\,b,\,c$ 
and $d$ 
\begin{eqnarray}
S(a,b,c,d)=\varepsilon^{\mu\nu\rho\sigma}a_\mu b_\nu c_\rho d_\sigma\,.
\end{eqnarray}
Requiring in addition gauge invariance
we were led to a restricted set of 13 gauge invariant amplitudes which 
are listed in Table \ref{tab1}. 
Then the invariant matrix element can 
be represented as a linear superposition of this set of basic 
amplitudes, i.e., 
\begin{eqnarray}
{\cal M}_{fi}=\sum_{\alpha=1}^{13} 
 F_\alpha(s,t,u)\, \Omega_\alpha\,, \label{mficov}
\end{eqnarray}
where $ F_\alpha(s,t,u)$ denote invariant functions which depend 
solely on the Mandelstam variables $s$, $t$ and $u$, defined as usual by
\begin{eqnarray}
s = (k+p)^2\,,\quad 
t = (k-q)^2\,,\quad
u = (k-p'\,)^2\,,
\end{eqnarray}
of which only two are independent, 
for example, $s$ and $t$. It is obvious that the specific form of the 
invariant functions $F_\alpha(s,t,u)$ will 
depend on the detailed dynamical properties of the target and the produced 
meson. 

The first nine amplitudes of Table \ref{tab1} 
are purely transverse in the c.m.\ frame of photon and target particle, 
and thus are well suited for describing 
photoproduction. However, the remaining four, which in addition are needed in 
electroproduction, contain besides charge and longitudinal 
current components also transverse current pieces. For this reason 
we had replaced in \cite{Are98} the last four amplitudes of Table \ref{tab1} 
by equivalent amplitudes $\widetilde \Omega_\alpha$ 
which are purely longitudinal in the c.m.\ frame. 
This could be achieved by using the following relations given in Eqs.\ 
(33) through (36) of \cite{Are98} which we repeat here in a more compact 
form because there some 
closing brackets were missing and, more importantly, the term containing 
$\Omega_e$ should not have been present in Eq.\ (34), 
\begin{eqnarray}
k\cdot p\, \Omega_f(k,x) - k_\mu^2\,\Omega_f(p,x)
&=&(k_\mu^2+k\cdot p)\, \Omega_h(k,x)
- k_\mu^2\,\Omega_h(q,x)\label{equiv1}\nonumber\\
&& -[k,p\,;k,q]\,\Omega_b(k,p,x) + [k,p\,;k,p]\,\Omega_b(k,q,x)\,,\\
k\cdot p\, \Omega_g(x,k) - k_\mu^2\,\Omega_g(x,p)
&=&k\cdot p\, \Omega_h(x,k)
- k_\mu^2\,\Omega_h(x,q) \nonumber\\
&& +[k,p\,;k,q]\,\Omega_c(k,p,x) - [k,p\,;k,p]\,\Omega_c(k,q,x)
\,,\label{equiv4}
\end{eqnarray}
where $x\in \{k,q\}$, and the symbol $[a,b\,;c,d]$ is defined by
\begin{eqnarray}
[a,b\,;c,d]:=(a\cdot c)(b\cdot d)-(a\cdot d)(b\cdot c)\,.\label{defklammer}
\end{eqnarray}

Therefore, we had replaced the last four amplitudes of Table \ref{tab1} by 
the equivalent ones ($\widetilde \Omega_{\alpha}$) as given on the left 
hand side of (\ref{equiv1}) and (\ref{equiv4}) to be used for the 
longitudinal contributions in electroproduction. They are related to the 
$\Omega_\alpha$ according to (\ref{equiv1}) and (\ref{equiv4}) by
\beqa
\widetilde \Omega_{10/11}&=&(k_\mu^2+k\cdot p)\, \Omega_{2/3}
- k_\mu^2\,\Omega_{4/5} -[k,p\,;k,q]\,\Omega_{6/7} 
+ [k,p\,;k,p]\,\Omega_{10/11}
\,,\label{equiv1a}\\
\widetilde \Omega_{12/13}&=&k\cdot p\, \Omega_{2/4}
+[k,p\,;k,q]\,\Omega_{8/9} - [k,p\,;k,p]\,\Omega_{12/13}
\,.\label{equiv4a}
\eeqa
Then the reaction 
amplitude of (\ref{mficov}) becomes
\begin{eqnarray}
{\cal M}_{fi}=\sum_{\alpha=1}^{13} 
\widetilde F_\alpha(s,t,u)\,\widetilde \Omega_\alpha\,, \label{mfinew}
\end{eqnarray}
where $\widetilde \Omega_\alpha= \Omega_\alpha$ for $\alpha=1,\dots,9$, and 
for $\alpha=10,\dots,13$ they are listed in Table \ref{tab1b}. Furthermore, 
the associated new invariant functions $\widetilde F_\alpha(s,t,u)$ are 
related to the $ F_\alpha(s,t,u)$ introduced previously by 
\begin{eqnarray}
\widetilde F_1 &=&  F_1 \,,\nonumber\\
\widetilde F_2 &=&  F_2 -\frac{1}{[k,p\,;k,p]}
\,((k_\mu^2 +k\cdot p) F_{10}-k\cdot p F_{12})\,,\nonumber\\
\widetilde F_3 &=&  F_3 -\frac{k_\mu^2 +k\cdot p}{[k,p\,;k,p]}\,
 F_{11}\,,\nonumber\\
\widetilde F_4 &=&  F_4 +\frac{1}{[k,p\,;k,p]}\,
(k_\mu^2F_{10}+k\cdot p F_{13})\,,\nonumber\\
\widetilde F_5 &=&  F_5 +\frac{k_\mu^2}{[k,p\,;k,p]}\, F_{11}
\,,\nonumber\\
\widetilde F_{6/7} &=&  F_{6/7} +\frac{[k,p\,;k,q]}{[k,p\,;k,p]}\, 
F_{10/11}
\,,\nonumber\\
\widetilde F_{8/9} &=&  F_{8/9} +\frac{[k,p\,;k,q]}{[k,p\,;k,p]}\, 
F_{12/13}\,,\nonumber\\
\widetilde F_{10/11} &=&\frac{1}{[k,p\,;k,p]}\, F_{10/11}
\,,\nonumber\\
\widetilde F_{12/13} &=&-\frac{1}{[k,p\,;k,p]}\, F_{12/13}\,.\label{FFtilde}
\end{eqnarray}
In these relations, one has to express the various kinematic factors in 
terms of the Mandelstam variables by using
\beqa
k\cdot p &=&\frac{1}{2}\,(s-M^2+K^2)\,,\\
k\cdot q &=&\frac{1}{2}\,(m^2-K^2-t)\,,\\
p\cdot q &=&\frac{1}{2}\,(s+t-M^2+K^2)\,,
\eeqa
with $K^2=-k_\mu^2$.
Since $[k,p\,;k,p]=-(K^2M^2+(k\cdot p)^2) <0$, no kinematic singularities are 
introduced by these transformations. 

For the study of the crossing properties of the $\widetilde F_\alpha$, 
the following inverse relations are useful
\begin{eqnarray}
F_2 &=&  \widetilde F_2 +(k_\mu^2 +k\cdot p) \widetilde F_{10}
+k\cdot p \widetilde F_{12}\,,\nonumber\\
F_3 &=&  \widetilde F_3 +(k_\mu^2 +k\cdot p) \widetilde  F_{11}\,,\nonumber\\
F_4 &=&  \widetilde F_4 +k_\mu^2\widetilde F_{10}-k\cdot p \widetilde F_{13}\,,\nonumber\\
F_5 &=&  \widetilde F_5 -k_\mu^2 \widetilde F_{11}
\,,\nonumber\\
F_{6/7} &=&  \widetilde F_{6/7} -[k,p\,;k,q]\, \widetilde F_{10/11}
\,,\nonumber\\
F_{8/9} &=&  \widetilde F_{8/9} +[k,p\,;k,q]\, \widetilde F_{12/13}
\,,\nonumber\\
F_{10/11} &=&[k,p\,;k,p] \widetilde F_{10/11}
\,,\nonumber\\
F_{12/13} &=&-[k,p\,;k,p] \widetilde F_{12/13}\,.\label{FtildeF}
\end{eqnarray}

Although these new amplitudes allow a 
clear separation of the invariant functions into longitudinal and transverse 
contributions, they also carry a disadvantage, namely a much more 
complicated crossing behaviour as is discussed in the next section.

\section{Crossing properties}\label{crossing}
Now we will study the crossing properties of the invariant functions 
with respect to the interchange of the initial with the final spin-one 
particle. To this end we will study the general structure of the reaction 
matrix shown in Fig.\ \ref{fig1} for the $s$- and $u$-channels in analogy 
to electromagnetic pion production on the nucleon \cite{Don72}. The 
reaction matrix is governed by the current operator
\beq
j^\nu_{\gamma D\, \Pi}(x)=\Phi^\dagger_{D,\,\mu}(x)\, J^{\mu\nu\rho}(x)\,
\Phi_{D,\,\rho}(x)\,\Phi_\Pi(x)\,.
\eeq
Here, $\Phi_\Pi(x)$ represents the usual field operator of a neutral 
pseudoscalar field and $\Phi^\mu_{D}(x)$ the one of a charged, 
massive vector field
\beq
\Phi^\mu_{D}(x)=N_\pi\,\sum_\lambda \int \frac{d^3p}{2\omega_p}\,
\Big(U^\mu(\vec p,\lambda)\,A_{D}(\vec p,\lambda)e^{-ip\cdot x}
+U^{\mu\ast}(\vec p,\lambda)\,B_{\bar D}^\dagger(\vec p,\lambda)
e^{ip\cdot x}\Big)\,,
\eeq
where $N_\pi$ is a normalization constant, $U^\mu(\vec p,\lambda)$ a  
polarization vector, $A_{D}(\vec p,\lambda)$ the anihilation operator of a
$D$-particle with momentum $\vec p$ and helicity $\lambda$, and 
$B_{\bar D}^\dagger(\vec p,\lambda)$ the creation operator 
of a corresponding antiparticle $\bar D$. These operators transform 
under charge conjugation
\beqa
C\,A_{D}(\vec p,\lambda)\,C^{-1} &=& B_{\bar D}(\vec p,\lambda)\,,\\
C\,B_{\bar D}^\dagger(\vec p,\lambda)\,C^{-1} &=& 
A_{D}^\dagger(\vec p,\lambda)\,,\\
C\,U^\mu(\vec p,\lambda)\,C^{-1} &=& U^{\mu T}(\vec p,\lambda)\,.
\eeqa
For the $s$-channel, i.e., pseudoscalar 
production on $D$, one obtains for the 
current matrix element
\beqa
\langle D(p',\lambda')\Pi(q)|\epsilon_\nu(k,\lambda_\gamma)\,&&
j^\nu_{\gamma D \,\Pi}(0)|D(p,\lambda)\gamma(k,\lambda_\gamma)\rangle
= \nonumber\\
&&
N_\pi^3 \sum_{\alpha=1}^{13}  F_\alpha(s,t,u)\,
U_{\mu}^{\dagger}(\vec p^{\,\prime},\lambda')\, 
 \epsilon_\nu(k,\lambda_\gamma)\,
{\cal O}^{\mu\nu\rho}_\alpha(k,p,q) \,U_\rho(\vec p,\lambda)\,.\label{schannel}
\eeqa
Similarly, one finds for the $u$-channel, i.e., $\Pi$ production on a $\bar D$
\beqa
\langle \bar D(-p,\lambda)\Pi(q)|\epsilon_\nu(k,\lambda_\gamma)&&\,
j^\nu_{\gamma D\, \Pi}(0)|\bar D(-p',\lambda')\gamma(k,\lambda_\gamma)\rangle
= \nonumber\\
&&\hspace*{-.5cm}
N_\pi^3 \sum_{\alpha=1}^{13} F_\alpha(s,t,u)\,
U_{\mu}^{T}(-\vec p^{\,\prime},\lambda')\, 
  \epsilon_\nu(k,\lambda_\gamma)\,
{\cal O}^{\mu\nu\rho}_\alpha(k,p,q) \,U_{\rho}^{\ast}(-\vec p,\lambda)\,.
\label{uchannel}
\eeqa
Now using invariance under charge conjugation of $\epsilon_\nu j^\nu$ 
and the transformations 
\beq
C|\gamma(k,\lambda_\gamma)\rangle=-|\gamma(k,\lambda_\gamma)\rangle\,,\quad 
C|D(p,\lambda)\rangle=|\bar D(p,\lambda)\rangle\,,\quad 
C|\Pi(q)\rangle=|\Pi(q)\rangle\,,
\eeq
one finds  
\beqa
\langle D(p',\lambda')\Pi(q)|\epsilon_\nu(k,\lambda_\gamma)\,&&
j^\nu_{\gamma D \Pi}(0)|D(p,\lambda)\gamma(k,\lambda_\gamma)\rangle
= \nonumber\\
&&
-\langle \bar D(p',\lambda')\Pi(q)|\epsilon_\nu(k,\lambda_\gamma)\,
j^\nu_{\gamma D \Pi}(0)|\bar D(p,\lambda)\gamma(k,\lambda_\gamma)\rangle\,.
\eeqa
Inserting the explicit expressions from (\ref{schannel}) and 
(\ref{uchannel}), one obtains the general crossing relation 
\beqa
\sum_{\alpha=1}^{13}  F_\alpha(s,t,u)\,&&
U_{\mu}^{\dagger}(\vec p^{\,\prime},\lambda')\, 
 \epsilon_\nu(k,\lambda_\gamma)\,
{\cal O}^{\mu\nu\rho}_\alpha(k,p,q) \,U_\rho(p,\lambda)=\nonumber\\
&&-\sum_{\alpha=1}^{13} F_\alpha(u,t,s)\,U_{\mu}^{T}(\vec p,\lambda)\, 
  \epsilon_\nu(k,\lambda_\gamma)\,
{\cal O}^{\mu\nu\rho}_\alpha(k,-p',q) \,U_{\rho}^{\ast}(p',\lambda')\,.
\label{gencross}
\eeqa
Introducing the transformation of the basic amplitudes $\Omega_\alpha$ under 
crossing, i.e., $U^\ast (\vec p^{\,\prime},\lambda')\leftrightarrow 
U(\vec p,\lambda)$ 
and $p \leftrightarrow -p'$ 
($s \leftrightarrow u$), according to
\beqa
\Omega_\alpha = U_{\mu}^{\dagger}(\vec p^{\,\prime},\lambda')\, 
\epsilon_\nu O^{\mu\nu\rho}_\alpha (k,p,q)\,U_\rho(\vec p,\lambda)&&\nonumber\\
\stackrel{\mbox{{\small crossing}}}{\longrightarrow}\quad
\Omega_\alpha^c &=& U_{\mu}^{T}(\vec p,\lambda)\, 
\epsilon_\nu O^{\mu\nu\rho}_\alpha (k,-p',q)
\,U_\rho^{\ast}(\vec p^{\,\prime},\lambda')\,,
\eeqa
the general crossing relation of (\ref{gencross}) then reads 
\beq
\sum_{\alpha=1}^{13} F_\alpha(s,t,u)\,\Omega_\alpha\,=
-\sum_{\alpha=1}^{13} F_\alpha(u,t,s)\,\Omega_\alpha^c
\,.\label{gencrossa}
\eeq

From the explicit expressions of the basic amplitudes in Table \ref{tab0}
one finds easily the following crossing transformations 
\begin{eqnarray}
\Omega_e(\epsilon,U',U) &\rightarrow & -\Omega_e(\epsilon,U',U)\,,\nonumber\\ 
\Omega_h(x,y) &\rightarrow & -\Omega_h(x,y)\,,\nonumber\\ 
\Omega_{b/c}(k,p,x) &\rightarrow & \Omega_{c/b}(k,p,x)
-\Omega_{c/b}(k,q,x)\,,\nonumber\\ 
\Omega_{b/c}(k,q,x) &\rightarrow & -\Omega_{c/b}(k,q,x)\,, 
\end{eqnarray}
for $x,y\in \{k,q\}$. This leads to the crossing transformations 
of the gauge invariant amplitudes of Table \ref{tab1}
\beqa
\Omega_{\alpha}^c &=& -\Omega_{\alpha}
\,,\mbox{ for }\,\alpha=1,2,5\,,\nonumber\\
\Omega_{3/4}^c &=& -\Omega_{4/3}\,,\nonumber\\
\Omega_{6/7}^c &=& \Omega_{8/9}-\Omega_{12/13}
\,,\nonumber\\
\Omega_{8/9}^c &=& \Omega_{6/7}-\Omega_{10/11}
\,,\nonumber\\
\Omega_{10/11}^c &=& -\Omega_{12/13}\,,\nonumber\\
\Omega_{12/13}^c &=& -\Omega_{10/11}\,.
\label{omcross}
\eeqa
Thus the crossing relation (\ref{gencrossa}) requires the invariant functions 
$F_{\alpha}(s,t,u)$ to possess the following crossing properties
\begin{eqnarray}
F_{\alpha}(s,t,u) &=& F_{\alpha}(u,t,s)\,,\mbox{ for }\,
\alpha=1,2,5\,,\nonumber\\
F_{3/4}(s,t,u) &=& F_{4/3}(u,t,s)\,,\nonumber\\
F_{6/7}(s,t,u) &=& -F_{8/9}(u,t,s)\,,\nonumber\\
F_{8/9}(s,t,u) &=& -F_{6/7}(u,t,s)\,,\nonumber\\
F_{10/11}(s,t,u) &=& F_{8/9}(u,t,s)+F_{12/13}(u,t,s)\,,\nonumber\\
F_{12/13}(s,t,u) &=& F_{6/7}(u,t,s)+F_{10/11}(u,t,s)\,.\label{Fcross}
\end{eqnarray}

For the alternative basic amplitudes $\widetilde \Omega_{\alpha}$ one 
finds for $\alpha=1,\dots,5$ the same crossing behaviour as for 
$\Omega_{\alpha}$ in (\ref{omcross}). For the other amplitudes one 
has to eliminate from the above crossing transformations the 
amplitudes $\Omega_{10/\dots/13}$ by using (\ref{equiv1a}) and 
(\ref{equiv4a}) resulting in  
\beqa
\widetilde \Omega_{6/7}^c &=& 
\frac{1}{[k,p\,;k,p]}\,(-k\cdot p \widetilde \Omega_{2/4} 
+[k,p\,;k,p']\widetilde \Omega_{8/9}-\widetilde \Omega_{12/13})
\,,\nonumber\\
\widetilde \Omega_{8/9}^c &=& 
\frac{1}{[k,p\,;k,p]}\,((k_\mu^2+k\cdot p) \widetilde \Omega_{2/3}
-k_\mu^2\,\widetilde \Omega_{4/5}+[k,p\,;k,p']\widetilde \Omega_{6/7}
-\widetilde \Omega_{10/11})
\,,\nonumber\\
\widetilde \Omega_{10/11}^c &=& 
\frac{1}{[k,p\,;k,p]}\,(-k_\mu^2\,[k,p\,;q,p]\widetilde \Omega_{2/4}
+k_\mu^2\,[k,p\,;k,p]\widetilde \Omega_{3/5}\nonumber\\
&&\hspace*{1.5cm}+((k\cdot p)^2-(k\cdot p')^2)
[k,p'\,;k,p']\widetilde \Omega_{8/9}+
[k,p'\,;k,p]\widetilde \Omega_{12/13})\,,\nonumber\\
\widetilde \Omega_{12/13}^c &=& 
\frac{1}{[k,p\,;k,p]}\,(k_\mu^2\,[k,p\,;(k+p),q]\widetilde \Omega_{2/3}
+k_\mu^2\,[k,p'\,;k,p]\widetilde \Omega_{4/5}\nonumber\\
&&\hspace*{1.5cm}-((k\cdot p)^2-(k\cdot p')^2)
[k,p'\,;k,p']\widetilde \Omega_{6/7}+
[k,p'\,;k,p]\widetilde \Omega_{10/11})\,.
\label{tomcross}
\eeqa
Correspondingly, the crossing properties of the invariant functions 
$\widetilde F_\alpha(s,t,u)$ are more involved. Explicitly one finds 
using either (\ref{tomcross}) or (\ref{Fcross}) in conjunction
with (\ref{FFtilde}) and (\ref{FtildeF})
\beqa
\widetilde F_1(s,t,u) &=&\widetilde F_1(u,t,s)\,,\nonumber\\
\widetilde F_2(s,t,u) &=&\widetilde F_2(u,t,s)+\frac{1}{[k,p\,;k,p]}(
k\cdot p\,\widetilde F_6(u,t,s)-(k_\mu^2 +k\cdot p)\,\widetilde F_8(u,t,s)
\nonumber\\
&&+k_\mu^2\,[k,p\,;q,p]\,\widetilde F_{10}(u,t,s)
-k_\mu^2\,[k,p\,;(k+p),q]\,\widetilde F_{12}(u,t,s))
\,,\nonumber\\
\widetilde F_3(s,t,u) &=&\widetilde F_4(u,t,s)-\frac{1}{[k,p\,;k,p]}(
(k_\mu^2 +k\cdot p)\,\widetilde F_9(u,t,s)\nonumber\\
&&
+k_\mu^2\,[k,p\,;k,p]\,\widetilde F_{10}(u,t,s)
+k_\mu^2\,[k,p\,;(k+p),q]\,\widetilde F_{13}(u,t,s))
\,,\nonumber\\
\widetilde F_4(s,t,u) &=&\widetilde F_3(u,t,s)+\frac{1}{[k,p\,;k,p]}(
k\cdot p\,\widetilde F_7(u,t,s)+k_\mu^2\,\widetilde F_8(u,t,s)\nonumber\\
&&
+k_\mu^2\,[k,p\,;q,p]\,\widetilde F_{11}(u,t,s)
-k_\mu^2\,[k,p'\,;k,p]\,\widetilde F_{12}(u,t,s))
\,,\nonumber\\
\widetilde F_5(s,t,u) &=&\widetilde F_5(u,t,s)-\frac{1}{[k,p\,;k,p]}(
k_\mu^2\,\widetilde F_9(u,t,s)\nonumber\\
&&
-k_\mu^2\,[k,p\,;k,p]\,\widetilde F_{11}(u,t,s)
-k_\mu^2\,[k,p'\,;k,p]\,\widetilde F_{13}(u,t,s))
\,,\nonumber\\
\widetilde F_{6/7}(s,t,u) &=&\frac{1}{[k,p\,;k,p]}(
-[k,p'\,;k,p]\,\widetilde F_{8/9}(u,t,s)\nonumber\\
&&
+((k\cdot p)^2-(k\cdot p')^2)\,[k,p'\,;k,p']\,\widetilde F_{12/13}(u,t,s))
\,,\nonumber\\
\widetilde F_{8/9}(s,t,u) &=&\frac{1}{[k,p\,;k,p]}(
-[k,p'\,;k,p]\,\widetilde F_{6/7}(u,t,s)\nonumber\\
&&
-((k\cdot p)^2-(k\cdot p')^2)\,[k,p'\,;k,p']\,\widetilde F_{10/11}(u,t,s))
\,,\nonumber\\
\widetilde F_{10/11}(s,t,u) &=&\frac{1}{[k,p\,;k,p]}(
\widetilde F_{8/9}(u,t,s)-[k,p'\,;k,p']\,\widetilde F_{12/13}(u,t,s))
\,,\nonumber\\
\widetilde F_{12/13}(s,t,u) &=&\frac{1}{[k,p\,;k,p]}(
\widetilde F_{6/7}(u,t,s)-[k,p'\,;k,p']\,\widetilde F_{10/11}(u,t,s))
\,.
\eeqa

If one wants simpler crossing properties of the basic amplitudes and 
invariant functions, one can introduce another alternative set of gauge 
invariant amplitudes $\widehat \Omega_\alpha$ by taking appropriate 
linear combinations of the original $\Omega_\alpha$
\beqa
\widehat \Omega_{\alpha}&=& \Omega_{\alpha}
\,,\mbox{ for }\,\alpha=1,2,5\,,\\
\widehat \Omega_{3/4}&=&\Omega_3 \pm \Omega_4\,,\\
\widehat \Omega_{6/8}&=&\Omega_6 \pm \Omega_8 
-\frac{1}{2}\widehat \Omega_{10/12}\,,\\
\widehat \Omega_{7/9}&=&\Omega_7 \pm \Omega_9
-\frac{1}{2}\widehat \Omega_{11/13}\,,\\
\widehat \Omega_{10/12}&=&\Omega_{10} \pm \Omega_{12}\,,\\
\widehat \Omega_{11/13}&=&\Omega_{11} \pm \Omega_{13}\,.
\eeqa
The explicit forms of those which are different from $\Omega_{\alpha}$ 
are listed in Table \ref{tab1c} where we have introduced in addition to the 
shorthand $[a,b\,;c,d]$ defined in (\ref{defklammer}) 
\beqa
\{a,b\,;c,d\}&:=&(a\cdot c)(b\cdot d)+(a\cdot d)(b\cdot c)\,,
\label{defantikom}\\
\widehat S^\mu(x)&:=&\varepsilon^{\mu\nu\rho\sigma} 
\epsilon_\nu k_\rho x_\sigma\,.\label{Sbar}
\eeqa
Then one may represent the invariant matrix element ${\cal M}_{fi}$ 
in terms of these new amplitudes introducing appropriate new invariant 
functions $\widehat F_\alpha(s,t,u)$ by 
\beq
{\cal M}_{fi}=\sum_{\alpha=1}^{13} 
\widehat F_\alpha(s,t,u)\,\widehat \Omega_\alpha\,. \label{mfigoodc}
\eeq
The $\widehat F_\alpha$ are related to the original invariant functions 
$F_\alpha$ by
\beqa
\widehat F_\alpha &=& F_\alpha,\mbox{ for }\, 
\alpha=1,2,5\,,
\nonumber \\
\widehat F_{3/4} &=& \frac{1}{2}\,(F_{3}\pm F_{4})\,,
\nonumber \\
\widehat F_{6/8} &=& \frac{1}{2}\,(F_{6}\pm F_{8})\,,
\nonumber \\
\widehat F_{7/9} &=& \frac{1}{2}\,(F_{7}\pm F_{9})\,,
\nonumber \\
\widehat F_{10/12} &=& \frac{1}{2}\,(F_{10}\pm F_{12} + 
\frac{1}{2}\,(F_{6}\pm F_{8}))\,,
\nonumber \\
\widehat F_{11/13} &=& \frac{1}{2}\,(F_{11}\pm F_{13} + 
\frac{1}{2}\,(F_{7}\pm F_{9}))\,.
\eeqa
The new amplitudes $\widehat \Omega_{\alpha}$ now 
have very simple crossing properties, namely
\beq
\widehat \Omega_{\alpha}^c=\left\{
\begin{array}{rl}
\widehat \Omega_{\alpha}, & \mbox{ for }\,\alpha=1,2,3,5,8,9,10,11,\\
-\widehat \Omega_{\alpha}, & \mbox{ for }\,\alpha=4,6,7,12,13,
\end{array}
\right.\label{hatcross}
\eeq
from which follows a correspondingly simple crossing behaviour of 
the $\widehat F_\alpha$ 
\beq
\widehat F_\alpha(s,t,u)=\left\{
\begin{array}{rl}
\widehat F_\alpha(u,t,s),& \mbox{ for }\,\alpha=4,6,7,12,13,\\
-\widehat F_\alpha(u,t,s),& \mbox{ for }\,\alpha=1,2,3,5,8,9,10,11.
\end{array}
\right.
\eeq
However, one has to keep in mind that 
these new amplitudes do not separate into transverse and 
longitudinal ones.

\section{Multipole decomposition}\label{multipole}

As in \cite{Are98}, we choose as basic reference frame one which has 
its $z$-axis parallel 
to the incoming real or virtual photon. For real photons the $x$-axis 
is chosen in the direction of maximal linear polarization, whereas 
for virtual photons the $y$-axis is taken perpendicular to the electron 
scattering plane, i.e., parallel to $\vec k_1 ^{lab}\times \vec k_2^{lab}$ 
where $k_1 ^{lab}$ and $k_2 ^{lab}$ denote the
lab frame momenta of the initial and the scattered electrons, respectively. 
The final state is described by the angles $\theta$ and $\phi$ of the 
momentum $\vec q$ with respect to the basic frame of reference.

Taking the representation of the 
$(S=1)$-spinors as given in (37) and (38) of \cite{Are98}, 
the helicity representation of the invariant amplitude can be written 
in the form
\beqa
T_{\lambda'\lambda_\gamma\lambda}(\theta,\phi)&=&
   e^{i(\lambda_\gamma -\lambda)\phi}
   t_{\lambda'\lambda_\gamma\lambda}(\theta)
\eeqa
with
\beqa
t_{\lambda'\lambda_\gamma\lambda}(\theta)&=& \sum_\alpha 
\widetilde F_\alpha (s,t,u)\,
\widetilde \Omega_{\alpha,\lambda'\lambda_\gamma\lambda}(\theta)
\,,\label{helrep}
\eeqa
thus separating the $\phi$-dependence. Here, 
$\widetilde F_\alpha$ and $\widetilde \Omega_\alpha$ may be replaced by 
$F_\alpha$ and $\Omega_\alpha$ or $\widehat F_\alpha$ and 
$\widehat \Omega_\alpha$ depending on the choice of basic amplitudes.
Explicit expressions of the helicity matrix elements of the various 
invariant amplitudes $\widetilde \Omega_\alpha$ 
are listed in Table 5 of \cite{Are98}. 
Note that in \cite{Are98} we had chosen $\phi=0$. 

The reduced 
matrix element $t_{\lambda'\lambda_\gamma\lambda}(\theta)$ will 
now be expanded into charge or longitudinal, electric and 
magnetic multipoles.
For the definition of the multipole decomposition we extend the 
convention of \cite{WiA95} for real photons to virtual ones including
charge multipoles as follows 
\begin{eqnarray}
t_{\lambda'\lambda_\gamma\lambda}(\theta)&=&
  \sqrt{\frac{1+\delta_{\lambda_\gamma 0}}{2}}\sum_L\sum_{lj}
\frac{\hat L \hat l}{\hat \jmath}(l01-\lambda'|j -\lambda')
(1 -\lambda L \lambda_\gamma|j\, \lambda_\gamma-\lambda)\,
O^{L \lambda_\gamma}_{l j}d^j_{\lambda_\gamma-\lambda,-\lambda'}(\theta)\,,
\end{eqnarray}
where $\hat \jmath=\sqrt{2j+1}$, and $(j_1m_1j_2m_2|jm)$ denotes a 
Clebsch-Gordan coefficient. 
Here $j$ and $l$ denote total and orbital angular momentum, respectively, 
of a final state partial wave, and $L$ the e.m.\ multipolarity. 
Furthermore,  $O^{L \lambda_\gamma}_{l j}$ contains the electric 
$(E^{L}_{l j})$, magnetic $(M^{L}_{l j})$
and coulomb or longitudinal $(C^{L}_{l j})$ multipoles according to 
\begin{eqnarray}
O^{L \lambda_\gamma}_{l j}=\delta_{|\lambda_\gamma|,1}
(E^{L}_{l j}+ \lambda_\gamma M^{L}_{l j}) + 
\delta_{\lambda_\gamma,0}C^{L}_{l j}\,.
\end{eqnarray}
Thus the various multipole types are obtained from 
$O^{L \lambda_\gamma}_{l j}$ by
\begin{eqnarray}
C^{L}_{lj}&=&O^{L 0}_{l j}\,,\label{cmult}\\
(E/M)^{L}_{l j}&=&\frac{1}{2}(O^{L 1}_{l j}\pm O^{L -1}_{l j})
\label{emmult}\,.
\end{eqnarray}
Using the orthogonality properties of the $d^j$-matrices \cite{Ros57}, 
one finds
\begin{eqnarray}
O^{L \lambda_\gamma}_{l j}&=&\frac{\hat \jmath}
{\sqrt{2(1+\delta_{\lambda_\gamma 0})}}
\sum_{\lambda'\lambda}\,C^{1lj}_{-\lambda',0,\lambda'}\,
C^{1Lj}_{\lambda,-\lambda_\gamma,\lambda_\gamma-\lambda}\,
\int_{-1}^1 d(\cos\theta)\,t_{\lambda'\lambda_\gamma\lambda}(\theta)\,
 d^j_{\lambda-\lambda_\gamma,\lambda'}(\theta)\,.
\end{eqnarray}
where we have defined 
\beq
C^{j_1 j_2 j_3}_{m_1,m_2,m_3}=
\left(\matrix{
j_1 & j_2 & j_3\cr
m_1 & m_2 & m_3\cr
}\right)\,
\eeq
as a shorthand for a $3j$-symbol. 
From (\ref{cmult}) and (\ref{emmult}) one obtains in detail
\beqa
C^{L}_{l j}&=&\frac{\hat \jmath}{2}
\sum_{\lambda'\lambda}
C^{1lj}_{-\lambda',0,\lambda'}\,
C^{1Lj}_{\lambda,0,-\lambda}\,
\int_{-1}^1 d(\cos\theta)\,t_{\lambda'0\lambda}(\theta)\,
 d^j_{\lambda,\lambda'}(\theta)\,,\label{cmulta}\\
(E/M)^{L}_{l j}&=&\frac{1}{2}\,(1\mp (-)^{l+L})\,
\frac{\hat \jmath}{\sqrt{2}}\,
\sum_{\lambda'\lambda}
\,C^{1lj}_{-\lambda',0,\lambda'}
\,C^{1Lj}_{\lambda,-1,1-\lambda}\,
\int_{-1}^1 d(\cos\theta)\,t_{\lambda'1\lambda}(\theta)\,
 d^j_{\lambda-1,\lambda'}(\theta)\label{emmulta}\,,
\eeqa
where in the latter case use has been made of the parity property
\beq
t_{-\lambda'-\lambda_\gamma-\lambda}=
(-)^{1+\lambda'+\lambda_\gamma+\lambda}
t_{\lambda'\lambda_\gamma\lambda}\,.
\eeq
The resulting multipole transitions, which can reach a given 
partial wave $(l,j)$, are listed in Table \ref{tabmult}. 

For the explicit representation of the multipoles one has to introduce 
the expansion of the reduced $t$-matrix into invariant amplitudes 
from (\ref{helrep}). Factorizing the helicity representation of 
the invariant amplitudes in a helicity independent part 
$\widetilde \rho^{\,\alpha}$ and a helicity dependent part 
$\widetilde \omega^{\,\alpha}_{\lambda' \lambda_\gamma \lambda}$
as listed in Table \ref{tabhelicity} by
\beq
\widetilde \Omega_{\alpha,\,\lambda' \lambda_\gamma \lambda}(\theta)=
\widetilde \rho^{\,\alpha}\,
\widetilde \omega^{\,\alpha}_{\lambda' \lambda_\gamma \lambda}(\theta)\,,
\eeq
and introducing
\beq
o^{(lj)L\lambda_\gamma}_\alpha(\theta) = 
\sum_{\lambda'\lambda}C^{1lj}_{-\lambda',0,\lambda'}
C^{1Lj}_{\lambda,-\lambda_\gamma,\lambda_\gamma-\lambda}\,
\widetilde \omega^{\,\alpha}_{\lambda' \lambda_\gamma \lambda}(\theta)\,
d^j_{\lambda-\lambda_\gamma,\lambda'}(\theta)\,,\label{oljL}
\eeq
one finds for the multipoles the compact form
\beq
O^{L \lambda_\gamma}_{l j}=\frac{\hat \jmath}
{\sqrt{2(1+\delta_{\lambda_\gamma 0})}}
\sum_{\alpha}\widetilde \rho^{\,\alpha}\,
\int_{-1}^1 d(\cos\theta)\,\widetilde F_\alpha(s,t,u)\,
o^{(lj)L\lambda_\gamma}_\alpha(\theta)\,,
\eeq
where for $|\lambda_\gamma|=1$ the sum over $\alpha$ runs from 1 through 9
and for $\lambda_\gamma=0$ from 10 through 13. The evaluation of the 
$o^{(lj)L\lambda_\gamma}_\alpha(\theta)$ is straightforward although 
somewhat lengthy. With the help of the addition theorem of the 
$d^j$-matrices \cite{Ros57}, 
one can expand the $o^{(lj)L\lambda_\gamma}_\alpha(\theta)$ 
in terms of Legendre polynomials.  
Explicit expressions are given in the Appendix A.

\section{Differential cross section and polarization observables}
\label{observable}
For the representation of observables 
we will take the same formal framework than used in deuteron photo- and 
electrodisintegration following the treatment in \cite{Are88,ArS90,ALT93}. 
Any observable in coherent scalar or pseudoscalar photo- or 
electroproduction off a spin-one target may be given in the form
\beq
{\cal O}_X =  3c\,\mbox{Tr}\,(T^\dagger \widehat O_X T\rho)\,,
\label{genobs}
\eeq
where the trace refers to the spin degrees of freedom of initial 
and final states, and $c$ is a kinematic factor explained below. 
Here $X$ characterizes an observable associated with the analysis 
of the final target spin state. We choose the representation 
$X= (IM\pm)$ $(I=0,1,2;\,I\ge M\ge 0)$ with a corresponding hermitean 
operator
\beq
\widehat O_{IM{\rm sig}_M}=c_{M{\rm sig}_M}(\tau^{[I]}_M 
+{\rm sig}_M (-)^M \tau^{[I]}_{-M})\,,
\eeq
with 
\beq
c_{M{\rm sig}_M}=\left\{\matrix{
\frac{1}{1+\delta_{M0}}, & \mbox{ for }\,{\rm sig}_M=+\,,\cr
i, & \mbox{ for }\,{\rm sig}_M= -\,.\cr}\right.
\eeq
Here we have introduced a sign function by ${\rm sig}_M:=\pm$, where 
the subscript $M$ merely indicates to which variable it refers. 

The irreducible tensors $\tau^{[I]}$ are the usual statistical tensors
for the parametrization of the density matrix of a spin-one particle
with normalization 
$\langle 1 ||\tau^{[I]}||1\rangle = \sqrt{3}\,\hat I$, i.e.\ in detail
\beq
\tau^{[I]} = \left\{
\begin{array}{cl}
 {\mathbb 1}_3 & \mbox{ no polarization,}\\
     \sqrt{\frac{3}{2}}\,S^{[1]} & \mbox{ vector polarization,}\\
     \sqrt{3}\,S^{[2]} & \mbox{ tensor polarization,}
\end{array}
\right.
\label{deuttensor}
\eeq
where ${\mathbb 1}_3$ is the unit matrix, $S^{[1]}$ the spin operator, 
and $S^{[2]}=[S^{[1]}\times S^{[1]}]^{[2]}$ the tensor operator of a
spin-one particle. In particular, we note the property
\beq
(\tau^{[I]}_M)^\dagger = (-)^M\,\tau^{[I]}_{-M}\,.\label{tauherm}
\eeq

Furthermore, the initial state density matrix $\rho$ is composed from 
the density matrices of the photon $\rho^\gamma$ and the one of the 
spin-one particle $\rho^D$ 
\beq
\rho = \rho^\gamma\otimes \rho^D\,.
\eeq
For virtual photons, i.e., electron scattering, the kinematic factor 
$c$ reads 
\begin{equation}
c(k_1^{lab},k^{lab}_2) = {\alpha \over 6 \pi^2} 
                          {k^{lab}_2 \over k_1^{lab} q_{\nu}^4}\;,
\end{equation}
where $\alpha$ is the fine structure constant and 
$q_{\nu} ^2=q_0^2-\vec q^{\,2}$ denotes the four momentum transfer 
squared $(q = k_1 - k_2)$.
In this case the density matrix is given by
\beq
\rho^{\gamma^*}_{\lambda_\gamma \lambda_\gamma'}=\sum_{\alpha= \{L,T,LT,TT\}}
(\delta^\alpha_{\lambda_\gamma \lambda_\gamma'}\rho_\alpha 
+h\delta^{\prime\alpha}_{\lambda_\gamma \lambda_\gamma'}\rho_\alpha')\,,
\label{virtualdm}
\eeq
where
\beqa
\rho_L = -\beta^2 q^2_{\nu} {\xi^2 \over 2 \eta}\,,\hspace{.9cm}& &\hspace{1cm}
\rho_{LT} = -\beta q^2_{\nu} {\xi \over \eta} \sqrt{{\xi + \eta \over 8}}\;,
\\
\rho_T = -{1 \over 2} q^2_{\nu} (1 + {\xi \over 2 \eta})\,,& &\hspace{1cm}
\rho_{TT} =  q^2_{\nu} {\xi \over 4 \eta}\;,\\
\rho'_{T} =- {1 \over 2} q^2_{\nu} \sqrt{{\xi + \eta \over \eta}}
\,,\,\,\,\,& &\hspace{1cm}
\rho'_{LT} = -{1 \over 2} \beta q^2_{\nu} {\xi \over \sqrt{2 \eta}}
\;,
\eeqa
with
\begin{equation}
\beta = {|{\vec q}^{\,lab}| \over |{\vec q}^{\,c}|},\,\,\,\,\,
\xi = -{q^2_{\nu} \over ({\vec q}^{\,lab})^{\,2}},\,\,\,\,\,
\eta = {\rm tan}^2({\theta_e^{lab} \over 2})\;.
\end{equation}
Here $\theta_e^{lab}$ denotes the electron scattering angle in the lab system 
and $\beta$ expresses the boost from the lab system to the frame in which the 
$T$-matrix is evaluated and ${\vec q}^{\,c}$ denotes the momentum 
transfer in this frame. 

For real photons only the transverse polarization components contribute and 
the corresponding density matrix for real photons is obtained from the 
one for virtual photons in (\ref{virtualdm}) by the replacements 
\beqa
\matrix{\rho_L \rightarrow 0,\,&\rho_{LT} \rightarrow 0,\,&
\rho'_{LT} \rightarrow 0,\,\cr
\rho_T \rightarrow \frac{1}{2},\,&
h\rho'_T \rightarrow \frac{1}{2}P^\gamma_c,\,& 
\rho_{TT} \rightarrow -\frac{1}{2}P^\gamma_l\,,\cr}
\eeqa
where $P^\gamma_l$ and $P^\gamma_c$ denote the degree of linear and circular
photon polarization, respectively. For the kinematic factor one has 
$c= 1/3$. 

As mentioned above, the density matrix of a spin-one particle is parametrized 
in terms of the irreducible tensors $\tau^{[I]}$ in spin-one space 
\beq
\rho^D = \frac{1}{3}\sum_{I=0}^2\sum_{M=-I}^I 
\tau^{[I]}_{M}\,P^{D\ast}_{IM}\,,
\eeq
where $P^{D}_{IM}$ characterizes the initial state polarization. In view of
the experimental methods for orienting deuterons, it is sufficient to 
assume that the 
density matrix is diagonal with respect to a certain orientation axis, 
characterized by spherical angles 
$\theta_D$ and $\phi_D$. Then one can write
\beq
P^{D}_{IM}=P_I^D e^{iM\phi_D}d^I_{M0}(\theta_D)\,,
\eeq
with the vector ($P_1^D$) and tensor ($P_2^D$) polarization parameters. 

Inserting all these various expressions, one arrives finally at
\beqa
{\cal O}_X&=&P_X S_0\nonumber\\
&=&c\, \sum _{I=0}^2 P_I^D \sum _{M=0}^I
 \Big\{\Big(\rho _L f_L^{IM}(X) + \rho_T f_T^{IM}(X) + 
\rho_{LT} {f}_{LT}^{IM+}(X) \cos \phi \nonumber\\
& &\qquad \qquad \qquad + \rho _{TT} {f}_{TT}^{IM+}(X) \cos2 \phi\Big)
\cos (M\tilde{\phi}-\bar\delta_{I}^{X} {\pi \over 2}) \nonumber\\
& &\;\;-\Big(\rho_{LT} {f}_{LT}^{IM-}(X) \sin \phi 
+ \rho _{TT} {f}_{TT}^{IM-}(X) \sin2 \phi\Big) 
\sin (M\tilde{\phi}-\bar\delta_{I}^{X} {\pi \over 2})\nonumber\\
& &\mbox{}+ h \Big[ \Big(\rho'_T f_T^{\prime\, IM}(X) 
+ \rho '_{LT} {f}_{LT}^{\prime\, IM-}(X) \cos \phi \Big)
\sin (M\tilde{\phi}-\bar\delta_{I}^{X} {\pi \over 2}) \nonumber\\
& &\qquad + \rho '_{LT} {f}_{LT}^{\prime\, IM+}(X) \sin \phi 
\cos (M\tilde{\phi}-\bar\delta_{I}^{X} {\pi \over 2})\Big] 
\Big\} d_{M0}^I(\theta_D)\,,\label{obsfin}
\eeqa
where $\tilde{\phi}=\phi-\phi_D$, and 
$S_0$ denotes the unpolarized differential cross section while  
$X=(I'M'{\rm sig}_{M'})$ characterizes the final polarization analysis of the 
spin-one particle. Furthermore, we have introduced
\beq
\bar \delta_I^X:= (\delta_{X,B}-\delta_{I1})^2\,,
\eeq
and 
\beq
\delta_{X,B}:=\left\{\begin{array}{ll}
0 & \mbox{for}\; X\in A:=
\{00+,\,11-,\, 20+,\, 21+,\,22+\} \\
 1 & \mbox{for}\; X\in B:=\{10+,\,11+,\, 21-,\,22-\} \\
\end{array}\right.\,,
\eeq
distinguishing two sets of recoil polarization observables $A$ and $B$ 
according to their transformation under parity.

The structure functions, which are the basic quantities 
containing the information on the dynamics of the reaction, are defined 
in complete analogy to \cite{ALT93} where 
we have expressed all structure functions in terms of real or 
imaginary parts of quantities ${\cal U}_{X}^{\lambda' \lambda I M}$ 
defined below. In detail one finds (note $M \ge 0$)
\beqa
f_{L}^{IM}(X)&=&\frac{2}{1+\delta_{M0}}
\Re e\left(i^{\bar \delta^X_I}
{\cal U}^{00 I M}_{X}\right)\,,\label{fL}\\
f_{T}^{IM}(X)&=&\frac{4}{1+\delta_{M0}}
\Re e\left(i^{\bar \delta^X_I}
{\cal U}^{11 I M}_{X}\right)\,,\\
f_{LT}^{IM{\rm sig}_M}(X)&=&\frac{4}{1+\delta_{M0}}
\Re e\left[i^{\bar \delta^X_I}
\left({\cal U}^{01 I M}_{X}+\mbox{sig}_M
(-)^{I+M+\delta_{X,\,B}}
{\cal U}^{01 I -M}_{X}\right)\right]\,,\\
f_{TT}^{IM{\rm sig}_M}(X)&=&\frac{2}{1+\delta_{M0}}
\Re e\left[i^{\bar \delta^X_I}
\left({\cal U}^{-11 I M}_{X}+\mbox{sig}_M
(-)^{I+M+\delta_{X,\,B}}
{\cal U}^{-11 I -M}_{X}\right)\right]\,,\\
f_{T}^{\prime IM}(X)&=&\frac{4}{1+\delta_{M0}}
\Re e\left(i^{1+\bar \delta^X_I}
{\cal U}^{11 I M}_{X}\right)\,,\\
f_{LT}^{\prime IM{\rm sig}_M}(X)&=&
\frac{4}{1+\delta_{M0}}\Re e
\left[i^{1+\bar \delta^X_I}
\left({\cal U}^{01 I M}_{X}+\mbox{sig}_M
(-)^{I+M+\delta_{X,\,B}}
{\cal U}^{01 I -M}_{X}\right)\right]\,,\label{fLTp}
\eeqa
where we have introduced for $M'\ge 0$ and $X=I'M'{\rm sig}_{M'}$
\beq
{\cal U}_{X}^{\lambda_\gamma' \lambda_\gamma I M}= 
c_{M'{\rm sig}_{M'}}
\Big({\cal U}_{I'M'}^{\lambda_\gamma' \lambda_\gamma I M}
+{{\rm sig}}_{M'}\,(-)^{M'}\,
{\cal U}_{I'-M'}^{\lambda_\gamma' \lambda_\gamma I M}\Big)
\eeq
with
\beq
{\cal U}_{I'M'}^{\lambda_\gamma' \lambda_\gamma I M}= 
\sum_{\bar \lambda' \lambda' \bar \lambda  \lambda}
t^*_{\bar \lambda' \lambda_\gamma'\bar \lambda}\langle \bar \lambda' |
\tau^{[I']}_{M'}|\lambda' \rangle
t_{\lambda'  \lambda_\gamma \lambda}
\langle \lambda|\tau^{[I]}_M|\bar \lambda\rangle
\,.\label{ulamcart}
\eeq
The latter quantities possess the following symmetries
\beqa
{\cal U}_{I'-M'}^{-\lambda_\gamma' -\lambda_\gamma I -M}&=&
(-)^{\lambda_\gamma'+\lambda_\gamma+I+M+I'+M'}
{\cal U}_{I'M'}^{\lambda_\gamma' \lambda_\gamma I M}\,,\\
\Big({\cal U}_{I'M'}^{\lambda_\gamma' \lambda_\gamma I M}\Big)^\ast&=&
(-)^{M+M'}{\cal U}_{I'-M'}^{\lambda_\gamma \lambda_\gamma' I -M}\,.
\eeqa
Combined, these then lead to the property
\beq
\Big({\cal U}_{X}^{\lambda_\gamma' \lambda_\gamma I M}
\Big)^\ast = (-)^{\lambda_\gamma'+ \lambda_\gamma +I +\delta_{X,B}}
{\cal U}_{X}^{-\lambda_\gamma -\lambda_\gamma' I M}\,,
\eeq
which has been used in the above representation of the structure functions.

As next we will evaluate the ${\cal U}$'s in terms of the invariant functions
$\widetilde F_\beta$. The helicity $t$-matrix element is given in terms of 
the helicity matrix elements of the basic amplitudes
\beq
t_{\lambda'  \lambda_\gamma \lambda}= \sum_\beta \widetilde F_\beta\,
\widetilde \Omega_{\beta,\lambda'  \lambda_\gamma \lambda}\,,
\eeq
where $\beta=1,\dots,9$ for $|\lambda_\gamma|=1$, and 
$\beta=10,\dots,13$ for $\lambda_\gamma=0$. They have the general form
\beq
\widetilde \Omega_{\beta,\lambda'  \lambda_\gamma \lambda}=
U^{\mu\dagger}(\vec p^{\,\prime},\lambda')\, 
\widetilde c^{\,\lambda_\gamma}_{\beta,\mu\nu}\,U^\nu(\vec p,\lambda)\,,
\label{omegalsl}
\eeq
where the quantities $\widetilde c^{\,\lambda_\gamma}_{\beta,\mu\nu}$ can be 
read off from the explicit expressions in Table \ref{tabhelicity}.  
They are listed in Table \ref{cmunu} and possess the symmetry
\beq
(\widetilde c^{\,\lambda_\gamma}_{\beta,\mu\nu})^\ast = (-)^{\lambda_\gamma}
\widetilde c^{\,-\lambda_\gamma}_{\beta,\mu\nu}\,.
\eeq
Inserting now these expressions into the defining equation (\ref{ulamcart}) 
for ${\cal U}_{I'M'}^{\lambda_\gamma' \lambda_\gamma I M}$, 
one obtains the representation in terms of the invariant functions 
\beq
{\cal U}_{I'M'}^{\lambda_\gamma' \lambda_\gamma I M}=
\sum_{\beta' \beta}\widetilde F_{\beta'}^\ast\,\widetilde F_\beta\,
\widetilde G^{\lambda_\gamma'\lambda_\gamma;\, \beta' \beta}_{I'M';\,I M}
\,,\label{UIM}
\eeq
where we have introduced for convenience
\beq
\widetilde G^{\lambda_\gamma'\lambda_\gamma;\,\beta' \beta}_{I'M';\,I M}=
(\widetilde c^{\,\lambda_\gamma'}_{\beta',\mu'\mu})^\ast\,
\widetilde c^{\,\lambda_\gamma}_{\beta,\nu'\nu}\,
{\cal G}^{\mu'\nu'}_{p',I'M'}\,{\cal G}^{\nu\mu}_{p,IM}\,,
\eeq
and
\beq
{\cal G}^{\mu\nu}_{p,IM}=
\sum_{\bar \lambda  \lambda}
U^\mu(\vec p,\bar \lambda)\,
\langle \bar \lambda|\tau^{[I]}_M|\lambda\rangle\,
U^{\nu\dagger}(\vec p,\lambda)\,.
\eeq
With the help of (\ref{tauherm}), one easily proves the property
\beq
({\cal G}^{\mu\nu}_{p,IM})^\ast=(-)^M\,{\cal G}^{\nu\mu}_{p,I-M}\,,
\eeq
from which follows
\beq
(\widetilde G^{\lambda_\gamma'\lambda_\gamma;
\,\beta' \beta}_{I'M';\,I M})^\ast=
(-)^{M'+M}\,\widetilde G^{\lambda_\gamma\lambda_\gamma';
\,\beta \beta'}_{I'-M';\,I -M}\,.\label{Gsymm}
\eeq
Details of 
the explicit evaluation of the $\widetilde G^{\lambda_\gamma'\lambda_\gamma;
\,\beta' \beta}_{I'M';\,I M}$ are given in the Appendix B. 

Eq.\ (\ref{UIM}) exhibits a clear separation into dynamical properties 
represented by the invariant functions $\widetilde F_\beta$ and kinematics 
described by the $\widetilde 
G^{\lambda_\gamma'\lambda_\gamma;\,\beta' \beta}_{I'M';\,I M}$ because 
the latter depend on kinematic quantities only, 
i.e., on the statistical tensors of initial and final deuteron polarization 
states and the basic amplitudes. 

With this representation of all observables in terms of the 
invariant functions $\widetilde F_\beta$ we will conclude the discussion 
of formal aspects of the invariant amplitudes for electromagnetic coherent 
pseudoscalar production on a spin-one target. It remains as a task for 
the future to derive explicitly the invariant functions associated with 
specific reaction diagrams, for example, the impulse approximation.

\setcounter{equation}{0}
\section*{Appendix A: Legendre polynomial expansion of the multipole
contributions of the basic amplitudes}

In this appendix, we list in Table \ref{tabolj} explicit expressions of the 
$o^{(lj)L\lambda_\gamma}_\alpha(\theta)$ as expansions in terms of Legendre
polynomials where we have introduced the following 
quantities $d^{(lj)L}_\beta$ which arise in evaluating the various helicity 
dependent terms of the 
$\widetilde \omega^\alpha_{\lambda'\lambda_\gamma\lambda}$ listed in 
Table \ref{tabhelicity}
\beqa
d^{(lj)L}_1(\theta)&=&\sum_{\lambda'\lambda}C^{1lj}_{-\lambda',0,\lambda'}
C^{1Lj}_{\lambda,-1,1-\lambda}
\,\delta_{\lambda' 0}\delta_{\lambda 0}\,d^1_{-1,0}(\theta)
\,d^j_{\lambda-1,\lambda'}(\theta)\nonumber\\
&=&-C^{1lj}_{0,0,0}\,
C^{1Lj}_{0,-1,1}\,\sum_J \hat{J}^2\,
C^{1jJ}_{0,0,0}\,C^{1jJ}_{-1,1,0}\,P_J(\cos\theta)\,,\\
d^{(lj)L}_2(E',\,E,\,\theta)&=&
\sum_{\lambda'\lambda}C^{1lj}_{-\lambda',0,\lambda'}
C^{1Lj}_{\lambda,-1,1-\lambda}
E'(\lambda')\,E(\lambda)\, d^1_{\lambda \lambda'}(\theta)
\,d^1_{-1 0}(\theta)\,d^1_{-1,0}(\theta)
\,d^j_{\lambda-1,\lambda'}(\theta)\nonumber\\
&=&(-)^{j+L}\,\sum_J \hat {J}^2\,
\Big[\sum_{J'} \hat{J'}^2\,G^{J'J}_{(lj)0}(E')\,G^{J'J}_{(Lj)1}(E)\Big]\,
P_J(\cos\theta)\,,\\
d^{(lj)L}_3(E',\,\theta)&=&
\sum_{\lambda'\lambda}C^{1lj}_{-\lambda',0,\lambda'}
C^{1Lj}_{\lambda,-1,1-\lambda}\,
E'(\lambda')\,\delta_{\lambda 0}\,d^1_{0\lambda'}(\theta)\,d^1_{-1,0}(\theta)
\,d^j_{\lambda-1,\lambda'}(\theta)\nonumber\\
&=&C^{1Lj}_{0,-1,1}\,\sum_J \hat{J}^2\,
\Big[\sum_{J'} \hat{J'}^2
\,C^{11J'}_{0,-1,1}\,C^{J'jJ}_{-1,1,0}
\,G^{J'J}_{(lj)0}(E')\Big]\,P_J(\cos\theta)\,,\\
d^{(lj)L}_4(E,\,\theta)&=&
\sum_{\lambda'\lambda}C^{1lj}_{-\lambda',0,\lambda'}
C^{1Lj}_{\lambda,-1,1-\lambda}\,
\delta_{\lambda' 0}\,E(\lambda)\, d^1_{\lambda 0}(\theta)\,d^1_{-1,0}(\theta)
\,d^j_{\lambda-1,\lambda'}(\theta)\nonumber\\
&=&(-)^{l+L}
C^{1lj}_{0,0,0}\,
\sum_J \hat{J}^2\,
\Big[\sum_{J'} \hat{J'}^2
\,C^{11J'}_{0,0,0}\,C^{J'jJ}_{0,0,0}
\,G^{J'J}_{(Lj)1}(E)\Big]\,P_J(\cos\theta)\,,\\
d^{(lj)L}_5(E',\,E,\,\theta)&=&
\sum_{\lambda'\lambda}C^{1lj}_{-\lambda',0,\lambda'}
C^{1Lj}_{\lambda,-1,1-\lambda}\,
E'(\lambda')\,E(\lambda)\,d^1_{0 \lambda'}(\theta)\,d^1_{\lambda 0}(\theta)
\,d^1_{-1,0}(\theta)
\,d^j_{\lambda-1,\lambda'}(\theta)\nonumber\\
&=&(-)^{j+L}\sum_J \hat{J}^2\,
\Big[\sum_{j',J'}  \hat{j'}^2 \hat{J'}^2\,(-)^{J'}\,
H^{(J'j')J}_{(lj)0}(E')\,H^{(J'j')J}_{(Lj)1}(E)\Big]\,
P_J(\cos\theta)\,,\\
d^{(lj)L}_6(E',\,\theta)&=&
\sum_{\lambda'\lambda}C^{1lj}_{-\lambda',0,\lambda'}
C^{1Lj}_{\lambda,-1,1-\lambda}\,
E'(\lambda')\,\delta_{\lambda 0}\,d^1_{-1,\lambda'}(\theta)
\,d^j_{\lambda-1,\lambda'}(\theta)\nonumber\\
&=&-C^{1Lj}_{0,1,-1}\,
\sum_J \hat{J}^2\,C^{1jJ}_{-1,1,0}\,D^J_{(lj)}(E')\,
P_J(\cos\theta)\,,\\
d^{(lj)L}_7(E',\,E,\,\theta)&=&
\sum_{\lambda'\lambda}C^{1lj}_{-\lambda',0,\lambda'}
C^{1Lj}_{\lambda,-1,1-\lambda}\,
E'(\lambda')\,E(\lambda)\, d^1_{\lambda 0}(\theta)
\, d^1_{-1 \lambda'}(\theta)\,d^j_{\lambda-1,\lambda'}(\theta)\nonumber\\
&=&(-)^{1+j+L}\sum_J \hat{J}^2\,
\Big[\sum_{J'} (-)^{J'} \hat{J'}^2\,G^{J'J}_{(lj)0}(E')\,
G^{J'J}_{(Lj)1}(E)\Big]\,
P_J(\cos\theta)\,,\\
d^{(lj)L}_8(\theta)&=&
\sum_{\lambda'\lambda}C^{1lj}_{-\lambda',0,\lambda'}
C^{1Lj}_{\lambda,-1,1-\lambda}\,
\delta_{\lambda' 0}\,\delta_{\lambda 1}\,
d^j_{\lambda-1,\lambda'}(\theta)\nonumber\\
&=&C^{1lj}_{0,0,0}\,C^{1Lj}_{1,-1,0}\,P_j(\cos\theta)\,,\\
d^{(lj)L}_{9}(E',\,\theta)&=&
\sum_{\lambda'\lambda}C^{1lj}_{-\lambda',0,\lambda'}
C^{1Lj}_{\lambda,-1,1-\lambda}\,
E'(\lambda')\, d^1_{0 \lambda'}(\theta)\,\delta_{\lambda 1}\,
d^j_{\lambda-1,\lambda'}(\theta)\nonumber\\
&=&C^{1Lj}_{1,-1,0}\,
\sum_J \hat{J}^2\,C^{1jJ}_{0,0,0}\,D^J_{(lj)}(E')\,
P_J(\cos\theta)\,,\\
d^{(lj)L}_{10}(\theta)&=&
\sum_{\lambda'\lambda}C^{1lj}_{-\lambda',0,\lambda'}
C^{1Lj}_{\lambda,0,-\lambda}\,\lambda'\, d^1_{\lambda' 0}(\theta)
\,\delta_{\lambda 0}\,d^j_{\lambda,\lambda'}(\theta)\nonumber\\
&=&(-)^{1+j+l}\,\frac{\sqrt{6}}{M}\,
C^{1Lj}_{0,0,0}\,
\sum_J \hat{J}^2\,C^{1jJ}_{0,0,0}\,G^{1J}_{(lj)0}(M)\,
P_J(\cos\theta)\,,\\
d^{(lj)L}_{11}(E,\,\theta)&=&
\sum_{\lambda'\lambda}C^{1lj}_{-\lambda',0,\lambda'}
C^{1Lj}_{\lambda,0,-\lambda}\,\lambda'\, d^1_{\lambda' 0}(\theta)
\,\lambda'\, d^1_{\lambda' 0}(\theta)\,E(\lambda)\, 
d^1_{\lambda 0}(\theta)\,d^j_{\lambda,\lambda'}(\theta)\nonumber\\
&=&(-)^{j+L}\,\frac{\sqrt{6}}{M}\,
\sum_J \hat{J}^2\,\Big[\sum_{J'}  \hat{J'}^2\,C^{1J'J}_{000}
\,D^{J'}_{(Lj)}(E)\,H^{(1J')J}_{(lj)0}(M)\Big]\,
P_J(\cos\theta)\,,\\
d^{(lj)L}_{12}(\theta)&=&
\sum_{\lambda'\lambda}C^{1lj}_{-\lambda',0,\lambda'}
C^{1Lj}_{\lambda,0,-\lambda}\,\delta_{\lambda' 0}\,\lambda\,
d^1_{\lambda 0}(\theta)\,d^j_{\lambda,\lambda'}(\theta)\nonumber\\
&=&(-)^{j+L}\,\frac{\sqrt{6}}{M}\,C^{1lj}_{000}
\sum_J \hat{J}^2\,C^{1jJ}_{0,0,0}\,G^{1J}_{(Lj)0}(M)\,
P_J(\cos\theta)\,,\\
d^{(lj)L}_{13}(E',\,\theta)&=&
\sum_{\lambda'\lambda}C^{1lj}_{-\lambda',0,\lambda'}
C^{1Lj}_{\lambda,0,-\lambda}\,E'(\lambda')
\, d^1_{0 \lambda'}(\theta)\,\lambda\,d^1_{\lambda 0}(\theta)\,
d^j_{\lambda,\lambda'}(\theta)\nonumber\\
&=&(-)^{L}\,\frac{\sqrt{6}}{M}\,
\sum_J \hat{J}^2\,\Big[\sum_{J'} (-)^{J'} \hat{J'}^2\,C^{1J'J}_{0,0,0}
\,D^{J'}_{(lj)}(E')\,H^{(1J')J}_{(Lj)0}(M)\Big]\,
P_J(\cos\theta)\,,
\eeqa
where $E'=E_q$ and $E=E_k$. Furthermore, we have defined for convenience 
\beqa
D^{J}_{(lj)}(e):&=&\sum_{\lambda}(-)^\lambda\,
C^{1jl}_{\lambda,-\lambda,0}\,C^{1jJ}_{\lambda,-\lambda,0}\,e(\lambda)\,,\\
G^{J'J}_{(Kj)m}(e):&=&\sum_{\lambda}\,
C^{1jK}_{\lambda,m-\lambda,-m}\,C^{11J'}_{\lambda,-m,m-\lambda}
\,C^{J'jJ}_{\lambda-m,m-\lambda,0}\,e(\lambda)\,,\\
H^{(J'j')J}_{(Kj)m}(e):&=&\sum_{\lambda}\,(-)^\lambda\,
C^{1jK}_{\lambda,m-\lambda,-m}\,C^{11J'}_{\lambda,0,-\lambda}
C^{1jj'}_{m,\lambda-m,-\lambda}
\,C^{J'j'J}_{\lambda,-\lambda,0}\,e(\lambda)\,.
\eeqa
In these expressions $e(\lambda)$ is defined as
\beqa
e(\lambda)&=& M\delta_{|\lambda|1} +e\delta_{\lambda 0}\nonumber\\
&=&(e+M)\delta_{\lambda 0}-(-)^\lambda M\,.
\eeqa
Inserting this, one finds with the help of sum rules
\beqa
D^{J}_{(lj)}(e):&=&(e+M)\,C^{1jl}_{0,0,0}\,C^{1jJ}_{0,0,0}
-\frac{M}{{\hat l}^2}\,\delta_{l,J}\,,\\
G^{J'J}_{(Kj)m}(e):&=&(e+M)\,C^{1jK}_{0,m,-m}\,C^{11J'}_{0,-m,m}
\,C^{J'jJ}_{-m,m,0}+(-)^{j}\,M\,C^{1KJ}_{m,-m,0}\,
\left\{\matrix{
J' & j & J \cr K & 1 & 1 \cr}\right\}\,,\\
H^{(J'j')J}_{(Kj)m}(e):&=&(e+M)\,
C^{1jK}_{0,m,-m}\,C^{11J'}_{0,0,0}\,
C^{1jj'}_{m,-m,0}\,C^{J'j'J}_{0,0,0}\nonumber\\
&&+(-)^{j+K}M\sum_{K'}{\hat{K'}}^2
\,C^{KJK'}_{m,0,-m}\,C^{K'11}_{-m,0,m}\left\{\matrix{
K & J & K' \cr 1 & J' & 1 \cr j & j' & 1 \cr}\right\}\,.
\eeqa

\setcounter{equation}{0}
\section*{Appendix B: Evaluation of the kinematic functions 
$\widetilde G^{\lambda_\gamma'\lambda_\gamma;
\,\beta' \beta}_{I'M';\,I M}$}

We will first evaluate ${\cal G}^{\mu\nu}_{p,IM}$ by 
using the explicit representation of the spin-one spinors
\beq
U^\mu(\vec p,\lambda)= (u_p)^\mu_{\,\,\nu}\,P^\nu\chi_{\lambda}^{[1]}\,,
\eeq
where the nonrelativistic spin-one spinor is 
denoted by $|\lambda \rangle=\chi_{\lambda}^{[1]}$, 
and the vector operator $P^\mu=(0,\vec P)$ 
has been introduced in Eq.\ (65) of \cite{Are98}. 
We note the orthogonality property 
\beq
P_kP_l^\dagger=\delta_{kl}\,,\label{pipk}
\eeq
and the relation 
\begin{eqnarray}
{P_k}^\dagger P_l=\frac{1}{3}\,\delta_{kl}\,{\mathbb 1}_3 +\frac{i}{2}\,
\varepsilon_{klj}\,S_j -S^{[2]}_{kl}\,,
\end{eqnarray}
from which follows 
\beq
\tau^{[I]}=- \sqrt{3}\,[P^{\dagger\,[1]}\times P^{[1]}]^{[I]}\,.\label{taui}
\eeq

Furthermore, one has
\beq
(u_p)^{\mu\nu}=g^{\mu\nu}-N_p\,\widetilde p^\mu\,\widetilde p^\nu
\eeq
with 
\beq
\widetilde p^\mu=M\delta^{\mu 0}+E_p\,p^\mu\quad \mbox{ and }\quad
N_p=\frac{1}{M(E_p+M)}\,.
\eeq
Then one finds in detail
\beqa
{\cal G}^{\mu\nu}_{p,IM}&=&{(u_p)^{\mu}}_{\mu'}\,
P^{\mu'} \tau^{[I]}_M P^{\nu'\dagger}\,{(u_p)_{\nu'}}^{\nu}\nonumber\\
&=&(\tau^{[I]}_M)^{\mu\nu}
-N_p\Big(\widetilde p^\mu (\tau^{[I]}_M)^{\nu\rho}\widetilde p_\rho
+\widetilde p_\rho (\tau^{[I]}_M)^{\rho\mu} \widetilde p^\nu\Big)
+N_p^2 \widetilde p^\mu \widetilde p^\nu (\tau^{[I]}_M)^{\rho\sigma}
\widetilde p_\rho \widetilde p_\sigma\,,\label{GpIM}
\eeqa
where we have defined 
\beq
(\tau^{[I]}_M)^{\mu\nu}= P^\mu \tau^{[I]}_M P^{\nu\dagger}\,.
\eeq
With the help of (\ref{taui}) and the orthogonality property (\ref{pipk})
one finds for the nonvanishing, i.e., purely spatial components in
spherical representation 
\beq
(\tau^{[I]}_M)_{\lambda' \lambda}=
(-)^{1+I}\sqrt{3}\,\widehat I\,C^{11I}_{\lambda',\lambda,M}\,.\label{taulsl}
\eeq
It possesses the symmetry
\beq
(\tau^{[I]}_M)^{\nu\mu}=(-)^{I}(\tau^{[I]}_M)^{\mu\nu}\,.\label{symtau}
\eeq
For convenience we introduce as abbreviations 
\beqa
\tau^{\mu\nu}_{IM}:&=&(\tau^{[I]}_M)^{\mu\nu}\,,\\
\tau^\mu_{IM;\,a}:&=&\tau^{\mu\nu}_{IM}a_\nu\,,\\
\,[ab]_{IM}:&=&a_\mu\,\tau^\mu_{IM;b}\,.
\eeqa
From the representation in (\ref{taulsl}) one easily finds 
\beq
\,[ab]_{IM}=-\sqrt{3}\,[a^{[1]}\times b^{[1]}]^{[I]}_M\,,
\eeq
in the notation of Fano-Racah for irreducible spherical tensors \cite{FaR59}. 
For later purpose we note according to (\ref{symtau}) the relation 
\beq
\tau^{\nu\mu}_{IM}a_\nu=(-)^{I}\tau^\mu_{IM;\,a}\,.
\eeq
In this notation, (\ref{GpIM}) reads
\beq
{\cal G}^{\mu\nu}_{p,IM}=\tau^{\mu\nu}_{IM}
-N_p\Big(\widetilde p^\mu\,\tau^\nu_{IM;\,p} 
+(-)^{I}\tau^\mu_{IM;\,p}\, \widetilde p^\nu\Big)
+N_p^2 \,\widetilde p^\mu \widetilde p^\nu\, [pp]_{IM}\,.\label{GpIMa}
\eeq
From (\ref{symtau}) follows furthermore 
\beq
{\cal G}^{\nu\mu}_{p,IM}=(-)^{I}{\cal G}^{\mu\nu}_{p,IM}\,,\label{symGmunu}
\eeq
since $[pp]_{IM}=0$ for $I=1$. 

For the evaluation of (\ref{UIM}) the following contractions are useful
\beqa
\tau^{\mu\nu}_{IM}\tau_{IM,\nu\mu}&=&(-)^M3\,
\delta_{I',\,I}\,\delta_{M',\,-M}
\,,\\
\,[ab]_{I'M';\,IM}:&=&\tau^\mu_{IM;\,a}\,g_{\mu\nu}\,\tau^\nu_{IM;\,b}
\nonumber\\
&=&-3\,\hat{I'} \hat I \sum_{JN} (-)^{J+N} \hat J \,C^{I'IJ}_{-M',-M,N}
\left\{ \matrix{
I' & I & J \cr
 1 & 1 & 1 \cr}\right\}
[a^{[1]}\times b^{[1]}]^{[J]}_N
\eeqa
The latter quantity obeys the symmetry
\beq
\,[ba]_{IM;\,I'M'}=(-)^{I'+I}[ab]_{I'M';\,IM}\,.
\eeq
In view of the explicit forms of 
$\widetilde c^{\,\lambda_\gamma}_{\beta,\mu\nu}$ in Table \ref{cmunu}, one 
needs for the evaluation of (\ref{UIM}) the following contractions 
introducing for convenience a compact notation
\beqa
{[}a\cdot{\cal G}_{p}\cdot b]_{IM}:&=&
a_\mu\,{\cal G}^{\mu\nu}_{p,IM}\,b_\nu\nonumber\\
&=&[ab]_{IM}
-N_p\Big(a\cdot\widetilde p\,[bp]_{IM} 
+(-)^{I}[ap]_{IM}\,b\cdot \widetilde p\Big)\nonumber\\
&&+N_p^2 \,a\cdot\widetilde p\, b\cdot\widetilde p \,[pp]_{IM}\,,\\
{[}{\cal G}_{p'}\cdot{\cal G}_{p}]_{I'M';\,IM}:&=&
{\cal G}^{\mu'\nu'}_{p',I'M'}\,g_{\mu'\mu}\,g_{\nu'\nu}\,
{\cal G}^{\nu\mu}_{p,IM}\nonumber\\
&=&
(-)^M3\,\delta_{I',\,I}\,\delta_{M',\,-M}\nonumber\\
&&-(1+(-)^{I'+I})\,\Big[N_{p'}\,[p'p']_{I'M';\,IM}+N_{p}\,[pp]_{I'M';\,IM}
\nonumber\\
&&+N_{p'}\,N_{p}\,([pp']_{I'M'}\,[p'p]_{IM}+(-)^{I'}\,
\widetilde p^{\,\prime}\cdot
\widetilde p\,[p'p]_{I'M';\,IM})\Big]
\nonumber\\
&&-N_{p'}^2\,N_{p}\,(1+(-)^{I})\,\widetilde p^{\,\prime}\cdot\widetilde p\,
[p'p']_{I'M'}\,[p'p]_{IM}
\nonumber\\
&&-N_{p'}\,N_{p}^2\,(1+(-)^{I'})\,\widetilde p^{\,\prime}\cdot\widetilde p\,
[pp']_{I'M'}\,[pp]_{IM}
\nonumber\\
&&+N_{p'}^2\,[p'p']_{I'M'}\,[p'p']_{IM}
+N_{p}^2\,[pp]_{I'M'}\,[pp]_{IM}
\nonumber\\
&&+N_{p'}^2\,N_{p}^2\,(\widetilde p^{\,\prime}\cdot\widetilde p)^2
[p'p']_{I'M'}\,[pp]_{IM}
\,,\\
{[}(a'\cdot{\cal G}_{p'})\cdot({\cal G}_{p}\cdot b)]_{I'M';\,IM}:&=&
a'_{\mu'}\,{\cal G}^{\mu'\nu'}_{p',I'M'}\,g_{\nu'\nu}\,
{\cal G}^{\nu\mu}_{p,IM}\,b_\mu\nonumber\\
&=&(-)^{I'}\,[a'b]_{I'M';\,IM}
\nonumber\\
&&-N_{p'}\,\Big(a'\cdot\widetilde p^{\,\prime}\,[p'b]_{I'M';\,IM}
+(-)^{I'+I}\,[a'p']_{I'M'}\,[bp']_{IM}\Big)
\nonumber\\
&&-N_{p}\,\Big((-)^{I'+I}\,[a'p]_{I'M';\,IM}\,b\cdot\widetilde p
+[a'p]_{I'M'}\,[bp]_{IM}\Big)
\nonumber\\
&&+N_{p'}\,N_{p}\,\Big(a'\cdot\widetilde p^{\,\prime}\,[pp']_{I'M'}\,[bp]_{IM}
+(-)^{I'}\,[a'p']_{I'M'}\,\widetilde p^{\,\prime}\cdot\widetilde p\,[bp]_{IM}
\nonumber\\
&&+(-)^{I}\,a'\cdot\widetilde p^{\,\prime}\,[p'p]_{I'M';\,IM}
\,b\cdot\widetilde p
+(-)^{I'+I}\,[a'p']_{I'M'}\,[p'p]_{IM}\,b\cdot\widetilde p\Big)
\nonumber\\
&&+N_{p'}^2\,a'\cdot\widetilde p^{\,\prime}\,[p'p']_{I'M'}\,[bp']_{IM}
+N_{p}^2\,b\cdot\widetilde p\,[pp]_{I'M'}\,[a'p]_{IM}
\nonumber\\
&&-N_{p'}^2\,N_{p}\,a'\cdot\widetilde p^{\,\prime}\,[p'p']_{I'M'}\,
\Big(\widetilde p^{\,\prime}\cdot\widetilde p\,[bp]_{IM}
+(-)^{I}\,[p'p]_{IM}\,b\cdot\widetilde p\Big)
\nonumber\\
&&-N_{p'}\,N_{p}^2\,b\cdot\widetilde p\,[pp]_{IM}\,
\Big(a'\cdot\widetilde p^{\,\prime}\,[pp']_{I'M'}
+(-)^{I'}\,[a'p']_{I'M'}\,
\widetilde p^{\,\prime}\cdot\widetilde p\Big)
\nonumber\\
&&+N_{p'}^2\,N_{p}^2\,a'\cdot\widetilde p^{\,\prime}\,
\widetilde p^{\,\prime}\cdot\widetilde p
\,b\cdot\widetilde p\,[p'p']_{I'M'}\,[pp]_{IM}
\,.
\eeqa
Using (\ref{symGmunu}) one finds
\beqa
{[}({\cal G}_{p'}\cdot a')\cdot(b\cdot {\cal G}_{p})]_{I'M';\,IM}:&=&
a'_{\nu'}\,{\cal G}^{\mu'\nu'}_{p',I'M'}\,g_{\mu'\mu}\,
{\cal G}^{\nu\mu}_{p,IM}\,b_\nu
\nonumber\\
&=&(-)^{I'+I}\,[(a'\cdot{\cal G}_{p'})\cdot({\cal G}_{p}\cdot b)]_{I'M';\,IM}
\,.
\eeqa

Now it is an easy task and straightforward 
to evaluate $\widetilde G^{\lambda_\gamma'\lambda_\gamma;
\,\beta' \beta}_{I'M';\,I M}$. As an example, we list here explicitly 
the longitudinal kinematic functions 
$\widetilde G^{00;\,\beta' \beta}_{I'M';\,I M}$ 
for $\beta'\leq\beta$ 
\beqa
\widetilde G^{00;\,10\,10}_{I'M';\,I M}
&=&c_{kp}^2[S\cdot{\cal G}_{p'}\cdot S]_{I'M'} 
[k\cdot{\cal G}_{p}\cdot k]_{IM}\,,\nonumber\\
\widetilde G^{00;\,10\,11}_{I'M';\,I M}
&=&c_{kp}^2[S\cdot{\cal G}_{p'}\cdot S]_{I'M'} 
[q\cdot{\cal G}_{p}\cdot k]_{IM}\,,
\nonumber\\
\widetilde G^{00;\,10\,12}_{I'M';\,I M}
&=&c_{kp}^2[S\cdot{\cal G}_{p'}\cdot k]_{I'M'} 
[S\cdot{\cal G}_{p}\cdot k]_{IM}\,,\nonumber\\
\widetilde G^{00;\,10\,13}_{I'M';\,I M}
&=&c_{kp}^2[S\cdot{\cal G}_{p'}\cdot q]_{I'M'} 
[S\cdot{\cal G}_{p}\cdot k]_{IM}\,,
\nonumber\\
\widetilde G^{00;\,11\,11}_{I'M';\,I M}
&=&c_{kp}^2[S\cdot{\cal G}_{p'}\cdot S]_{I'M'} 
[q\cdot{\cal G}_{p}\cdot q]_{IM}\,,\nonumber\\
\widetilde G^{00;\,11\,12}_{I'M';\,I M}
&=&c_{kp}^2[S\cdot{\cal G}_{p'}\cdot k]_{I'M'} 
[S\cdot{\cal G}_{p}\cdot q]_{IM}\,,
\nonumber\\
\widetilde G^{00;\,11\,13}_{I'M';\,I M}
&=&c_{kp}^2[S\cdot{\cal G}_{p'}\cdot q]_{I'M'} 
[S\cdot{\cal G}_{p}\cdot q]_{IM}\,,\nonumber\\
\widetilde G^{00;\,12\,12}_{I'M';\,I M}
&=&c_{kp}^2[k\cdot{\cal G}_{p'}\cdot k]_{I'M'} 
[S\cdot{\cal G}_{p}\cdot S]_{IM}\,,
\nonumber\\
\widetilde G^{00;\,12\,13}_{I'M';\,I M}
&=&c_{kp}^2[k\cdot{\cal G}_{p'}\cdot q]_{I'M'} 
[S\cdot{\cal G}_{p}\cdot S]_{IM}\,,\nonumber\\
\widetilde G^{00;\,13\,13}_{I'M';\,I M}
&=&c_{kp}^2[q\cdot{\cal G}_{p'}\cdot q]_{I'M'} 
[S\cdot{\cal G}_{p}\cdot S]_{IM}\,,
\eeqa
with $c_{kp}=K^2\,\epsilon(0)\cdot p$ and 
$S_\mu:=\varepsilon_{\mu\nu\rho\sigma}k^\nu p^\rho q^\sigma$, 
while for $\beta'>\beta$ they can be obtained from the symmetry relation in 
(\ref{Gsymm}).

\begin{table}[h]
\caption{Set of basic types of invariant amplitudes with 
$x,y,z\in \{k,p,q\}$ and $a,b,c \in \{U',\epsilon, U\}$.} 

\begin{tabular}{cc}
  notation & explicit form\\
\tableline
 $\Omega_a(x)$ & $S(U',\epsilon, U, x)$\\
 $\Omega_b(x, y, z)$ & $S(U',\epsilon, x, y ) U\cdot z$\\
 $\Omega_c(x, y, z)$ & $U'\cdot z\,S(\epsilon, U, x, y)$\\
 $\Omega_d(x, y, z)$ & $S(U', U, x, y )\epsilon \cdot z$\\
 $\Omega_e(a, b, c)$ & $S(a, k,p,q)\,b\cdot c$\\
 $\Omega_f(x, y)$ & $S(U',k, p, q)\,\epsilon \cdot x \,U\cdot y$\\
 $\Omega_g(x, y)$ & $U'\cdot x \,S(U, k, p, q)\,\epsilon \cdot y$\\
 $\Omega_h(x, y)$ & $U'\cdot x \, S(\epsilon,k, p, q)\,U\cdot y$\\
\end{tabular}
\label{tab0}
\end{table}

\begin{table}[h]
\caption{Set of independent gauge invariant amplitudes $\Omega_\alpha$.}

\begin{tabular}{ccc}
$\alpha$ & notation & explicit form\\
\tableline
1 & $\Omega_e(\epsilon,U',U)$ & $S(\epsilon,k,p,q)U'\cdot U$\\
2 & $\Omega_h(k,k)$ & $U'\cdot k\,S(\epsilon, k,p,q )U\cdot k$\\
3 & $\Omega_h(k,q)$ & $U'\cdot k\,S(\epsilon, k,p,q )U\cdot q$\\
4 & $\Omega_h(q,k)$ & $U'\cdot q\,S(\epsilon, k,p,q )U\cdot k$\\
5 & $\Omega_h(q,q)$ & $U'\cdot q\,S(\epsilon, k,p,q )U\cdot q$\\
6 & $\Omega_b(k,p,k)$ & $S(U',\epsilon, k,p ) U\cdot k$\\
7 & $\Omega_b(k,p,q)$ & $S(U',\epsilon, k,p ) U\cdot q$\\
8 & $\Omega_c(k,p,k)$ & $U'\cdot k\,S(\epsilon,U, k,p )$\\
9 & $\Omega_c(k,p,q)$ & $U'\cdot q\,S(\epsilon,U, k,p )$\\
\tableline
10 & $\Omega_b(k,q,k)$ & $S(U',\epsilon, k,q ) U\cdot k$\\
11 & $\Omega_b(k,q,q)$ & $S(U',\epsilon, k,q ) U\cdot q$\\
12 & $\Omega_c(k,q,k)$ & $U'\cdot k\,S(\epsilon,U, k,q )$\\
13 & $\Omega_c(k,q,q)$ & $U'\cdot q\,S(\epsilon,U, k,q )$\\
\end{tabular}
\label{tab1}
\end{table}

\begin{table}[h]
\caption{Equivalent set of independent gauge invariant longitudinal
amplitudes $\widetilde \Omega_\alpha$ for electroproduction.}

\begin{tabular}{ccc}
$\alpha$ & notation & explicit form\\
\tableline
10 & $k\cdot p\, \Omega_f(k,k) - k^2\,\Omega_f(p,k)$ 
& $S(U',k,p,q )\,U \cdot k\,
[k,\epsilon\,;p,k]$\\
11 & $k\cdot p\, \Omega_f(k,q) - k^2\,\Omega_f(p,q)$ 
& $S(U',k,p,q )\,U \cdot q\,
[k,\epsilon\,;p,k]$\\
12 & $k\cdot p\, \Omega_g(k,k) - k^2\,\Omega_g(k,p)$ & 
$U'\cdot k\,S(U, k,p,q )
[k,\epsilon\,;p,k]$\\
13 & $k\cdot p\, \Omega_g(q,k) - k^2\,\Omega_g(q,p)$ & 
$U'\cdot q\,S(U, k,p,q )
[k,\epsilon\,;p,k]$\\
\end{tabular}
\label{tab1b}
\end{table}

\begin{table}[h]
\caption{Alternative set of independent gauge invariant amplitudes 
$\widehat\Omega_\alpha$ with simple crossing properties, where 
$P=(p+p')/2$ and $\widehat S(x)$ is defined in (\protect{\ref{Sbar}}).}

\begin{tabular}{ccc}
$\alpha$ & basic amplitudes & explicit form\\
\tableline
3 & $\Omega_h(k,q)+\Omega_h(q,k)$ & $S(\epsilon, k,p,q )\{U',U\,;k,q\}$\\
4 & $\Omega_h(k,q)-\Omega_h(q,k)$ & $S(\epsilon, k,p,q )[U',U\,;k,q]$\\
6 & $\Omega_b(k,P,k)+\Omega_c(k,P,k)$ & $[U',U\,;\widehat S(P),k]$\\
7 & $\Omega_b(k,P,q)+\Omega_c(k,P,q)$ & $[U',U\,;\widehat S(P),q]$\\
8 & $\Omega_c(k,P,k)-\Omega_c(k,P,k)$ & $\{U',U\,;\widehat S(P),k\}$\\
9 & $\Omega_c(k,P,q)-\Omega_c(k,P,q)$ & $\{U',U\,;\widehat S(P),q\}$\\
\tableline
10 & $\Omega_b(k,q,k)+\Omega_c(k,q,k)$ & $[U',U\,;\widehat S(q),k]$\\
11 & $\Omega_b(k,q,q)+\Omega_c(k,q,q)$ & $[U',U\,;\widehat S(q),q]$\\
12 & $\Omega_b(k,q,k)-\Omega_c(k,q,k)$ & $\{U',U\,;\widehat S(q),k\}$\\
13 & $\Omega_b(k,q,q)-\Omega_c(k,q,q)$ & $\{U',U\,;\widehat S(q),q\}$\\
\end{tabular}
\label{tab1c}
\end{table}

\begin{table}[h]
\caption{Selection rules for multipole transitions of order $L$ to 
a given partial wave $(l,j)$.} 

\begin{tabular}{ccc}
  $l$ & $L$ (charge/electric) & $L$ (magnetic) \\
\tableline
$j-1$ & $j$ & $j-1$, $j+1$\\
$j$ & $j-1$, $j+1$ & $j$ \\
$j+1$ & $j$ & $j-1$, $j+1$\\
\end{tabular}
\label{tabmult}
\end{table}

\begin{table}[h]
\caption{Representation of the helicity matrix elements of the 
invariant amplitudes in factorized form 
$\widetilde \Omega_{\alpha,\,\lambda' \lambda_\gamma \lambda}(\theta)=
\widetilde \rho^{\,\alpha}\,\widetilde 
\omega^{\,\alpha}_{\lambda' \lambda_\gamma \lambda}(\theta)$.}
\begin{center}

\begin{tabular}{rll}
$\alpha$ & $\widetilde \rho^{\,\alpha}$ & 
$\widetilde \omega^{\,\alpha}_{\lambda' \lambda_\gamma \lambda}$\\
\tableline
1 & $\frac{i}{M^2}\,kq\sqrt{s}$ & $\lambda_\gamma\,
      [p'p\,\delta_{\lambda' 0}\delta_{\lambda 0}
      - E'(\lambda')E(\lambda)\, d^1_{\lambda \lambda'}(\theta)]
      \,d^1_{-\lambda_\gamma 0}(\theta)$\\
2 & $\frac{i}{M^2}\,k^2qs $ & $\lambda_\gamma\,
     [qk_0\,\delta_{\lambda' 0}+kE'(\lambda')\,d^1_{0\lambda'}(\theta)]
      \,d^1_{-\lambda_\gamma 0}(\theta)\delta_{\lambda 0}$\\[1.ex]
3 & $\frac{i}{M^2}\,kq\sqrt{s} $ & $\lambda_\gamma\,
     [qk_0\,\delta_{\lambda' 0}+kE'(\lambda')\,d^1_{0\lambda'}(\theta)]
      [kq_0\,\delta_{\lambda 0}+qE(\lambda)\,d^1_{\lambda 0}(\theta)]
      \,d^1_{-\lambda_\gamma 0}(\theta)$\\
4 & $\frac{i}{M^2}\,k^2q^2s^{3/2} $ & $\lambda_\gamma
      \,\delta_{\lambda' 0}\delta_{\lambda 0}
      \,d^1_{-\lambda_\gamma 0}(\theta)$\\
5 & $\frac{i}{M^2}\,k q^2s $ & $\lambda_\gamma\,\delta_{\lambda' 0}\,
      [kq_0\,\delta_{\lambda 0}+qE(\lambda)\,d^1_{\lambda 0}(\theta)]
      \,d^1_{-\lambda_\gamma 0}(\theta)$\\
6 & $\frac{i}{M^2}\,k^2s$ & $\lambda_\gamma\, E'(\lambda')
      \,\delta_{\lambda 0}\,d^1_{-\lambda_\gamma,\lambda'}(\theta)$\\
7 & $\frac{i}{M^2}\,k\sqrt{s}$ & $\lambda_\gamma\,E'(\lambda')
     \,[kq_0\,\delta_{\lambda 0}+qE(\lambda)\,d^1_{\lambda 0}(\theta)]
     \,d^1_{-\lambda_\gamma,\lambda'}(\theta)$\\
8 & $\frac{i}{M}\,k\sqrt{s}$ & $\lambda_\gamma\,
     [qk_0\,\delta_{\lambda' 0}+kE'(\lambda')\,d^1_{0\lambda'}(\theta)]
      \,\delta_{\lambda \lambda_\gamma}$\\
9 & $\frac{i}{M}\,kqs$ & $\lambda_\gamma\,\delta_{\lambda' 0}
     \,\delta_{\lambda_\gamma \lambda}$\\
\tableline
10 & $\frac{i}{M}\,\sqrt{K^2}\,k^3qs^{3/2} $ & $\lambda'\,
     d^1_{\lambda'0}(\theta)\,\delta_{ \lambda_\gamma 0}
     \,\delta_{\lambda 0}$\\
11 & $\frac{i}{M}\,\sqrt{K^2}\,k^2qs $ & $\lambda'\,d^1_{\lambda'0}(\theta)
     \,\delta_{ \lambda_\gamma 0}\,
     [kq_0\,\delta_{\lambda 0}+qE(\lambda)\,d^1_{\lambda 0}(\theta)]$
\\
12 & $-\frac{i}{M}\,\sqrt{K^2}\,k^2qs $ & $\lambda\,
     [qk_0\,\delta_{\lambda' 0}+kE'(\lambda')\,d^1_{0\lambda'}(\theta)]
      \,\delta_{ \lambda_\gamma 0}\,d^1_{\lambda0}(\theta) $\\
13 & $-\frac{i}{M}\,\sqrt{K^2}\,k^2q^2s^{3/2} $ & $\lambda\,
     \delta_{\lambda'0}\,\delta_{ \lambda_\gamma 0}
     \,d^1_{\lambda 0}(\theta)$\\
\end{tabular}
\end{center}
\label{tabhelicity}
\end{table}

\begin{table}[h]
\caption{Explicit expressions for the quantities 
$\widetilde c^{\,\lambda_\gamma}_{\beta,\mu\nu}$ defined in (\ref{omegalsl}) 
for transverse ($|\lambda_\gamma|=1$ 
and $\beta=1,\dots,9$) and longitudinal ($\lambda_\gamma=0$ and 
$\beta=10,\dots,13$) contributions with $c_{kp}=K^2\,\epsilon(0)\cdot p$,
$S^{(\lambda_\gamma)}:=S(\epsilon_{\lambda_\gamma},k,p,q)$ and 
$S_\mu:=\varepsilon_{\mu\nu\rho\sigma}k^\nu p^\rho q^\sigma$.}

\begin{tabular}{cccccc}
$\beta$ & explicit form & $\beta$ & explicit form & $\beta$ & explicit form \\
\tableline
1 & $g_{\mu\nu}S^{(\lambda_\gamma)}$ & 6 & $S_\mu k_\nu$ & 10 &
$S_\mu k_\nu c_{kp}$\\
2 & $k_{\mu}k_{\nu}S^{(\lambda_\gamma)}$ & 7 & $S_\mu q_\nu$ & 11 &
$S_\mu q_\nu c_{kp}$\\
3 & $k_{\mu}q_{\nu}S^{(\lambda_\gamma)}$ & 8 & $-k_\mu S_\nu $ & 
12 & $k_\mu S_\nu c_{kp}$\\
4 & $q_{\mu}k_{\nu}S^{(\lambda_\gamma)}$ & 9 & $-q_\mu S_\nu $ & 
13 & $q_\mu S_\nu c_{kp}$\\
5 & $q_{\mu}q_{\nu}S^{(\lambda_\gamma)}$ & & & & \\
\end{tabular}
\label{cmunu}
\end{table}

\begin{table}[h]
\caption{Explicit expressions for $o^{(lj)L\lambda_\gamma}_\alpha(\theta)$ 
defined in (\ref{oljL}) with $\lambda_\gamma=1$ for 
$\alpha=1,\dots,9$ and $\lambda_\gamma=0$ for 
$\alpha=10,\dots,13$.}
\begin{center}

\begin{tabular}{rl}
$\alpha$ & $o^{(lj)L\lambda_\gamma}_\alpha(\theta)$ \\
\tableline
1 & $p'p\,d^{(lj)L}_1(\theta)-d^{(lj)L}_2(E_q,\,E_k,\,\theta) $\\
2 & $k_0q\,d^{(lj)L}_1(\theta)+k\,d^{(lj)L}_3(E_q,\,\theta)$\\
3 & $k_0kq_0q\,d^{(lj)L}_1(\theta)+k^2q_0\,d^{(lj)L}_3(E_q,\,\theta)
+k_0q^2\,d^{(lj)L}_4(E_k,\,\theta)+kq\,
d^{(lj)L}_5(E_q,\,E_k,\,\theta)$\\
4 & $d^{(lj)L}_1(\theta)$\\
5 & $kq_0\,d^{(lj)L}_1(\theta)+q\,d^{(lj)L}_4(E_k,\,\theta)$\\
6 & $d^{(lj)L}_6(E_q,\,\theta)$\\
7 & $kq_0\,d^{(lj)L}_6(E_q,\,\theta)+q\,d^{(lj)L}_7(E_q,\,E_k,\,\theta)$\\
8 & $k_0q\,d^{(lj)L}_8(\theta)+k\,d^{(lj)L}_{9}(E_q,\,\theta)$\\
9 & $d^{(lj)L}_8(\theta)$\\
\tableline
10 & $d^{(lj)L}_{10}(\theta)$\\
11 & $kq_0\,d^{(lj)L}_{10}(\theta)+q\,d^{(lj)L}_{11}(E_k,\,\theta)$\\
12 & $k_0q\,d^{(lj)L}_{12}(\theta)+k\,d^{(lj)L}_{13}(E_q,\,\theta)$\\
13 & $d^{(lj)L}_{12}(\theta)$\\
\end{tabular}
\end{center}
\label{tabolj}
\end{table}

\end{document}